\documentclass[ms]{aastex}
\usepackage{spr-astr-addons}
\usepackage{url}

\RequirePackage{color}

\def\arcsec{\hbox{$^{\prime\prime}$}}

\renewcommand{\vec}[1]{\mbox{\boldmath $#1$}}

\newcommand*\Del{\mathrm{\Delta}}
\newcommand*\del{\mathrm{\delta}}

\newcommand{\uvec}[1]{\hat{\vec #1}}

\newcommand{\F}{{\ \mathrm F} }
\newcommand{\C}{{\ \mathrm C} }
\newcommand{\m}{{\ \mathrm m} }
\newcommand{\nm}{{\ \mathrm {nm}} }
\newcommand{\fmeter}{{\ \mathrm {fm}} }
\newcommand{\pmeter}{{\ \mathrm {pm}} }
\newcommand{\mm}{{\ \mathrm {mm}} }
\newcommand{\ua}{{\ \mathrm {ua}} }
\newcommand{\s}{{\ \mathrm s} }
\newcommand{\kg}{{\ \mathrm {kg}} }
\newcommand{\GeV}{{\ \mathrm {GeV}} }
\newcommand{\Hz}{{\ \mathrm {Hz}} }
\newcommand{\grad}{{\bf \nabla } }
\newcommand{\J}{{\ \mathrm J} }
\newcommand{\Ph}{{\it \Phi}}

\newcommand{\emaila}{wilhelm@mps.mpg.de}
\newcommand{\emailb}{bnd.app@iitbhu.ac.in}

\begin{document}

\title{Impact models of gravitational and electrostatic forces\\
{\large Potential energies, atomic clocks, gravitational anomalies and redshift}}

\shorttitle{Impact models of gravitation and electrostatics}
\shortauthors{K. Wilhelm and B.N. Dwivedi}
\author{Klaus Wilhelm}
\affil{Max-Planck-Institut f\"ur Son\-nen\-sy\-stem\-for\-schung
(MPS), 37077~G\"ottingen, Germany \\ \emaila}
\and
\author{Bhola N. Dwivedi}
\affil{Department of Physics, Indian Institute of Technology
(Banaras Hindu University), Varanasi-221005, India \\
\emailb}

Last updated on \today

\begin{abstract}
The far-reaching gravitational force is described by a heuristic impact model
with hypothetical massless entities propagating at the speed of light in
vacuum and transferring momentum and energy between massive bodies through
interactions on a local basis. In the original publication \citep{WiWiDw},
a spherical symmetric emission of secondary entities had been postulated.
The potential energy problems in gravitationally and electrostatically bound
two-body systems have been studied in the framework of this impact model of
gravity and of a proposed impact model of the electrostatic force
\citep{WiDwWi}. These studies have indicated that an anti-parallel emission of
a secondary entity\,--\,now called \emph{graviton}\,--\,with respect to the
incoming one is more appropriate. This article is based on the latter choice
and presents the modifications resulting from this change. The model has been
applied to multiple interactions of gravitons in large mass conglomerations in
several publications. They will be summarized here taking the modified
interaction process into account. In addition, the speed of photons as a
function of the gravitational potential are considered in this context
together with the dependence of atomic clocks and the redshift on the
gravitational potential.
\end{abstract}

%%%%%%%%%%%%%%%%%%%%%%%%%%%%%%%%%%%%%%%%%%%%%%%%%%%%%%%%%%%%%%%%%%%%%%%%%%%%%%
%%%%%%%%%%%%%%%%%%%%%%%%%%%%%%%%%%%%%%%%%%%%%%%%%%%%%%%%%%%%%%%%%%%%%%%%%%%%%%
%%%%%%%%%%%%%%%%%%%%%%%%%%%%%%%%%%%%%%%%%%%%%%%%%%%%%%%%%%%%%%%%%%%%%%%%%%%%%%
%%%%%%%%%%%%%%%%%%%%%%%%%%%%%%%%%%%%%%%%%%%%%%%%%%%%%%%%%%%%%%%%%%%%%%%%%%%%%%



%%%%%%%%%%%%%%%%%%%%%%%%%%%%%%%%%%%%%%%%%%%%%%%%%%%%%%%%%%%%%%%%%%%%%%%%%%%%%%
%%%%%%%%%%%%%%%%%%%%%%%%%%%%%%%%%%%%%%%%%%%%%%%%%%%%%%%%%%%%%%%%%%%%%%%%%%%%%%
%%%%%%%%%%%%%%%%%%%%%%%%%%%%%%%%%%%%%%%%%%%%%%%%%%%%%%%%%%%%%%%%%%%%%%%%%%%%%%
%%%%%%%%%%%%%%%%%%%%%%%%%%%%%%%%%%%%%%%%%%%%%%%%%%%%%%%%%%%%%%%%%%%%%%%%%%%%%%

\section{Introduction}
\label{s.intro}
%% Sect. 1

Newton's law of gravity gives the attraction between two spherical
symmetric bodies~A and B with masses~$M$ and $m$, a separation
distance, $r$ (large compared to the sizes of the particles), and at rest in
an inertial frame of reference. The force acting on~B is
%
%% Eq. 1
\begin{equation}
\vec K_{\rm G}(r) =
-\frac{G_{\rm N}\,M\,\uvec r}{r^2}\,m  ~ ,% Newton
\label{eq.Newton}
\end{equation}
where $G_{\rm N} = 6.674\,08(31) \times 10^{-11}\m^3\kg^{-1}\s^{-2}$ is the
constant of gravity\footnote{This value and those of other constants are taken
from CODATA 2014 \citep{Mohetal}.\label{footnote_CODATA}}, $\uvec r$ is the
unit vector of the radius vector $\vec r$ with origin at A, and $r = |\vec r|$.
The first term on the right-hand side represents the classical gravitational
field of the mass $M$.

In close analogy, Coulomb's law yields the force of the electrostatic
interaction between particles~C with charge~$Q$ and D with charge~$q$:
%
%% Eq. 2
\begin{equation}
\vec K_{\rm E}(r) =
\frac{Q\,\uvec r}{4\,\pi\,r^2\,\varepsilon_0}~q ~ ,% Coulomb
\label{eq.Coulomb}
\end{equation}
where $\varepsilon_0 = 8.854\,187\,817\,...~\times~10^{-12}\F\m^{-1}$ is the
electric constant in vacuum. Here charges
with opposite sign lead to attraction and with equal signs to repulsion.
%
% Eq. 3
\begin{equation}
\vec E_Q(r) = \frac{Q\,\uvec r}{4\,\pi\,r^2\,\varepsilon_0}% ~E\_field
\label{eq.E_field}
\end{equation}
is the classical electric field of a charge~$Q$.

For two electrons, for instance, the ratio is
%
%% Eq. 4
\begin{equation}
R^{\rm E}_{\rm G} = \frac{|\vec K_{\rm E}(r)|}
{|\vec K_{\rm G}(r)|} =  4.16574 \times 10^{42} ~.% ratio
\label{eq.ratio}
\end{equation}

Newton's law of gravitation yields a very good approximation of gravitational
forces, unless effects treated in the General Theory of Relativity (GTR)
\citep{Ein16}
are of importance.

Nevertheless, the physical processes\,--\,in particular the potential
energies\,--\,of the gravitational
and the electrostatic fields are still a matter of debate:\\
\citet{Pla09} wondered about the energy and momentum of the
electromagnetic field. A critique of the classical field theory by
\citet{WheFey} concluded that a theory of action at a distance,
originally proposed by \citet{Sch03}, avoids the direct notion of fields.
\citet[][p.\,231]{Lan01} calls the fact ``remarkable'' that the motion of a
closed system in response to external forces is determined by the same law as
its constituents. In this context, it should be recalled that \citet{Lau11}
considered radiation confined in a certain volume (,,Hohlraumstrahlung'') and
showed that the radiation contributed to the mass of the system according to
Einstein's mass-energy equation, see Eq.\,(\ref{eq.energy}).
In a discussion of energy-momentum conservation for gravitational fields,
\citet[][p.\,468]{Pen06} finds even in closed systems ``something a little
`miraculous' about how things all fit together, ... ''; and
\citet{Car98} wrote: ``... after all, potential energy is a rather mysterious
quantity to begin with ...''.

Related to the potential energy problem is the disagreement of
\citet{Woletal} and
\citet{Mueetal} on whether the frequency of an atomic clock\,--\,causing the
gravitational redshift\,--\,is sensitive to the gravitational potential
%
%% Eq. 5
\begin{equation}
U_{\rm G}(r) = -\frac{G_{\rm N}\,M}{r}% ~G\_potential
\label{eq.G_potential}
\end{equation}
or to the local gravity field $\vec{g} = \grad U$.

These remarks and disputes motivated us to think about electrostatic and
gravitational fields and the problems related to the potential energies.

%%%%%%%%%%%%%%%%%%%%%%%%%%%%%%%%%%%%%%%%%%%%%%%%%%%%%%%%%%%%%%%%%%%%%%%%%%%%%%
%%%%%%%%%%%%%%%%%%%%%%%%%%%%%%%%%%%%%%%%%%%%%%%%%%%%%%%%%%%%%%%%%%%%%%%%%%%%%%
%%%%%%%%%%%%%%%%%%%%%%%%%%%%%%%%%%%%%%%%%%%%%%%%%%%%%%%%%%%%%%%%%%%%%%%%%%%%%%
%%%%%%%%%%%%%%%%%%%%%%%%%%%%%%%%%%%%%%%%%%%%%%%%%%%%%%%%%%%%%%%%%%%%%%%%%%%%%%

\section{Gravitational and electrostatic interactions}
\label{s.grav_elect}
%% Sect. 2

If far-reaching fields have to be avoided, gravitational and electrostatic
models come to mind similar to the emission of photons from a radiation source
and their absorption or scattering somewhere else\,--\,thereby transferring
energy and momentum with the speed of light, $c_0 = 299\,792\,458\m\s^{-1}$,
in vacuum \citep{Poi00,Ein17,Com23,Lew26}.

A heuristic model of Newton's law of gravitation has been proposed by
\citet[][Paper\,1]{WiWiDw},--\,without far-reaching
gravitational fields\,--\,involving hypothetical massless
entities. Originally they had been called quadrupoles, but will be
called \emph{gravitons} now. In subsequent studies, conducted to test the model
hypothesis, it became evident that the energy and momentum could not be
conserved in a closed system without modifying the interaction process of the
gravitons with massive bodies and massless particles, such as photons.
The modification and the
consequences in the context of the gravitational potential energy will be
discussed in the following sections together with related topics.

The analogy between Newton's and Coulomb's laws suggest that in the
latter case an impact model might be appropriate as well\,--\,with
electric dipole entities transferring momentum and energy.
This has been proposed in Paper~2 \citep{WiDwWi}.
The equations governing the behaviour of gravitons and dipoles in the next
sections are very similar in line with the similarity of Newton's and
Coulomb's laws.

Both concepts are required for a description
of the gravitational redshift in terms of physical processes in
Sect.\,\ref{ss.redshift}.

%%%%%%%%%%%%%%%%%%%%%%%%%%%%%%%%%%%%%%%%%%%%%%%%%%%%%%%%%%%%%%%%%%%%%%%%%%%%%%
%%%%%%%%%%%%%%%%%%%%%%%%%%%%%%%%%%%%%%%%%%%%%%%%%%%%%%%%%%%%%%%%%%%%%%%%%%%%%%

\subsection{Definitions of gravitons}
\label{ss.gravitons}
%% Subsect. 2.1

Without a far-reaching gravitational field, the interactions have to be
understood on a local basis with energy and momentum transfer by gravitons.
This interpretation has several features in common with a theory
based on gravitational shielding conceived by Nicolas \citet{Fat90} at the
end of the seventeenth century. A French manuscript can be found in
\citet{Bop29}, and an outline in German has been provided by \citet{Zeh83}.
Related ideas by Le Sage have been discussed by \citet{Dru97}.

The gravitational case, in contrast to the electrostatic one, does not depend
on polarized particles. Gravitons with an electric
quadrupole configuration propagating with the speed of light~$c_0$
will be postulated in the case of gravity.
They are the obvious candidates as they have small interaction energies
with positive and negative electric charges, and, in addition, can easily be
constructed with a spin of $S = \pm\,2$, if indications to that effect are
taken into account
\citep[cf.][]{Wei64}.

The vacuum is thought to be permeated by the gravitons that are, in the
absence of near masses, isotropically distributed with (almost) no interaction
among each other\,--\,even dipoles have no mean interaction energy in the
classical theory \citep[see, e.g.][]{Jac99,Jac06}.
The graviton distribution is assumed to be a nearly stable, possibly
slowly varying quantity in space and time.
It has a constant spatial number density
%
%% Eq. 6
\begin{equation}
\rho_{\rm G} = \frac{\Del N_{\rm G}}{\Del V}~.% G\_density
\label{eq.G_density}
\end{equation}
Constraints on the energy spectrum of the gravitons will be considered in
later sections. At this stage we define a mean energy of
%
%% Eq. 7
\begin{equation}
T_{\rm G} = |\vec{p}_{\rm G}|\,c_0 = p_{\rm G}\,c_0% ~G\_energy
\label{eq.G_energy}
\end{equation}
for a massless graviton with a momentum vector of $\vec{p}_{\rm G}$.

%%%%%%%%%%%%%%%%%%%%%%%%%%%%%%%%%%%%%%%%%%%%%%%%%%%%%%%%%%%%%%%%%%%%%%%%%%%%%%
%%%%%%%%%%%%%%%%%%%%%%%%%%%%%%%%%%%%%%%%%%%%%%%%%%%%%%%%%%%%%%%%%%%%%%%%%%%%%%

\subsection{Definitions of dipoles}
\label{ss.dipoles}
%% Subsect. 2.2

A model for the electrostatic force can be obtained by introducing
hypothetical electric dipoles propagating with the speed of light.
The force is described by the action of dipole distributions on charged
particles. The dipoles are transferring momentum and energy between charges
through interactions on a local basis.

Apart from the requirement that the absolute values of the positive and
negative charges must be equal, nothing is known,
at this stage, about the values
themselves, so charges of~$\pm q$ will be assumed, where $q$ might or
might not be identical to the elementary charge
$e = 1.602\,176\,6208(98) \times 10^{-19}\C$.

The electric dipole moment is
%
% Eq. 8
\begin{equation}
\vec d = |q|\,\vec l% ~D\_moment
\label{eq.D_moment}
\end{equation}
parallel or antiparallel to the velocity vector $c_0\,\uvec n$, where
$\uvec n$ is a unit vector pointing in a certain direction.
This assumption is necessary in order to get attraction and repulsion
of charges depending on their mutual polarities. In Sect.~\ref{ss.Coulomb} it
will be shown that the value~$|\vec{d}|$ of the dipole moment is not critical
in the context of our model.
The dipoles have a mean energy
%
% Eq. 9
\begin{equation}
T_{\rm D} = |\vec{p_{\rm D}}|\,c_0 = p_{\rm D}\,c_0~ ,% D\_energy
\label{eq.D_energy}
\end{equation}
where $\vec{p_{\rm D}}$ represents the momentum of the dipoles.
As a working hypothesis, it will first be assumed that
$|\vec{p}_{\rm D}|$ is constant remote from gravitational centres with the
same value for all dipoles of an isotropic distribution.
%Later it will, however, be necessary to consider the
%energy spectrum of the dipoles.
The dipole distribution is assumed to be nearly stable in space and time with
a spatial number density
%
%% Eq. 10
\begin{equation}
\rho_{\rm D} = \frac{\Del N_{\rm D}}{\Del V}~, % D\_density
\label{eq.D_density}
\end{equation}
but will be polarized near electric charges and affected by
gravitational centres.

%%%%%%%%%%%%%%%%%%%%%%%%%%%%%%%%%%%%%%%%%%%%%%%%%%%%%%%%%%%%%%%%%%%%%%%%%%%%%%
%%%%%%%%%%%%%%%%%%%%%%%%%%%%%%%%%%%%%%%%%%%%%%%%%%%%%%%%%%%%%%%%%%%%%%%%%%%%%%

\subsection{Virtual entities}
\label{ss.virtual}
%% Subsect. 2.3

As an important step, a formal way will be outlined of achieving the required
momentum and energy transfers by discrete interactions. The idea is based on
virtual gravitons or dipoles in analogy with other virtual particles
\citep[cf., e.g.][]{Wei34,Yuk35,NimSta}.

%%%%%%%%%%%%%%%%%%%%%%%%%%%%%%%%%%%%%%%%%%%%%%%%%%%%%%%%%%%%%%%%%%%%%%%%%%%%%%

\subsubsection{Virtual gravitons}
\label{sss.vir_grav}
%% Subsect. 2.3.1

The virtual gravitons with energies of $T^*_{\rm G} \ll M\,c^2_0$ supplied by
a central body with mass~$M$ will have a certain lifetime $\Del t_{\rm G}$
and interact with ``real'' gravitons.
In the literature, there are many different derivations of an energy-time
relation \citep[cf.][]{ManTam,AhaBoh,Hil98}.
Considerations of the spread of the frequencies of a limited wave-packet
led \citet{Boh49} to an approximation for the indeterminacy of the energy
that can be re-written as
%
%% Eq. 11
\begin{equation}
T^*_{\rm G} \approx \frac{h}{\Del t_{\rm G}}% ~G\_lifetime
\label{eq.G_lifetime}
\end{equation}
with $h = 6.626\,070\,040(81) \times 10^{-34}\J\s$, the Planck constant.
For propagating gravitons, the equation
%
%% Eq. 12
\begin{equation}
T_{\rm G} = h\,\frac{c_0}{l_{\rm G}}% ~G\_Broglie
\label{eq.G_Broglie}
\end{equation}
is equivalent to the photon energy relation
$E_\nu = h\,\nu = h\,c_0/\lambda$, where
$\lambda$ corresponds to $l_{\rm G}$,
which can be considered as the wavelength of the
hypothetical gravitons.
Since there is experimental evidence that virtual photons (identified as
evanescent electromagnetic modes) behave non-locally
\citep{LowMen,StaNim},
the virtual gravitons might also behave non-locally.
Consequently, the absorption of a
real graviton could occur momentarily by a recombination with an appropriate
virtual one.

%%%%%%%%%%%%%%%%%%%%%%%%%%%%%%%%%%%%%%%%%%%%%%%%%%%%%%%%%%%%%%%%%%%%%%%%%%%%%%

\subsubsection{Virtual dipoles}
\label{sss.vir_dip}
%% Subsect. 2.3.2

\begin{figure}[t]
\begin{center}
\includegraphics[width=\columnwidth]{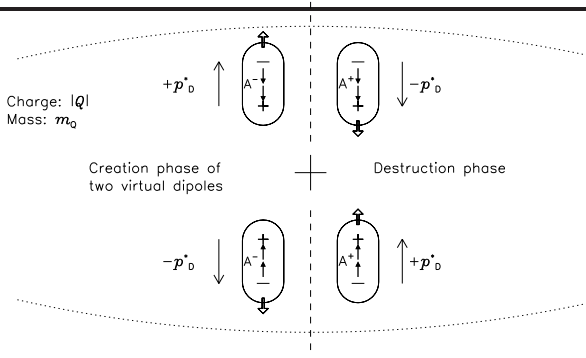}
\end{center}
\caption{Conceptional presentation of the creation and destruction phases
of virtual dipole pairs by a charge, and the corresponding momentum vectors
of the virtual dipoles (long arrows).
The dipoles are assumed to have a spin of
$S = \pm 1$~$(\pm 2 \times \hbar/2)$ (short arrows). (Figure~2 of Paper~2)
\label{fig:Creation}}
\end{figure}

We assume that a particle with charge~$Q$ is symmetrically
emitting virtual dipoles with $\vec p^*_{\rm D} $. The emission rate is
proportional to its charge, and the orientation such that a repulsion
exists between the charge and the dipoles. The symmetric creation and
destruction of virtual dipoles is sketched in Fig.~\ref{fig:Creation}.
The momentum balance is shown for the emission phase on the left and the
absorption phase on the right side.

%
%% ONE-COLUMN FIGURE
% Figure 1

Virtual dipoles with energies of $T^*_{\rm D} \ll m_Q\,c^2_0$ will have a
certain lifetime $\Del t_{\rm D}$ and interact with \emph{real} dipoles.
The momentum and energy relations correspond to those of the gravitons:
%
%% Eq. 13
\begin{equation}
T^*_{\rm D} \approx \frac{h}{\Del t_{\rm D}}~.% D\_lifetime
\label{eq.D_lifetime}
\end{equation}
The equation
%
%% Eq. 14
\begin{equation}
T_{\rm D} = h\,\frac{c_0}{l_{\rm D}}% ~D\_Broglie
\label{eq.D_Broglie}
\end{equation}
is also equivalent to the photon energy relation for propagating dipoles,
with $l_{\rm D}$ corresponding to $\lambda$.

%%%%%%%%%%%%%%%%%%%%%%%%%%%%%%%%%%%%%%%%%%%%%%%%%%%%%%%%%%%%%%%%%%%%%%%%%%%%%%
%%%%%%%%%%%%%%%%%%%%%%%%%%%%%%%%%%%%%%%%%%%%%%%%%%%%%%%%%%%%%%%%%%%%%%%%%%%%%%

\subsection{Newton's law of gravity}
\label{ss.Newton}
%% Subsect. 2.4

The gravitons are absorbed by massive bodies from the background and
subsequently emitted at rates determined by the mass~$M$ of the body
independent of its charge:
%
%% Eq.15
\begin{equation}
\frac{\Del N_M}{\Del t} = \rho_{\rm G}\,\kappa_{\rm G}\,M = \eta_{\rm G}\,M~ ,
%G\_absorption
\label{eq.G_absorption}
\end{equation}
where $\kappa_{\rm G}$ is the gravitational absorption
coefficient and $\eta_{\rm G}$ the corresponding emission coefficient.

\begin{figure}[t]
\begin{center}
\includegraphics[width=\columnwidth]{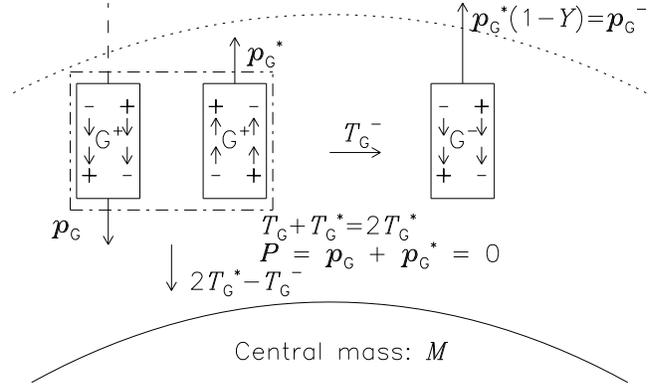}
\end{center}
\caption{Interaction of gravitons with a body of mass, $M$. A graviton
arriving with a momentum $\vec{p}_{\rm G}$ on the left combines
together with a virtual graviton with $\vec{p}_{\rm G}$$^* = -\vec{p}_{\rm G}$.
The excess energy liberates a second virtual graviton with
$\vec{p}^-_{\rm G}$ on the right in a
direction anti-parallel to the incoming graviton. The excess energy
$T^-_{\rm G}$ is smaller than $T^*_{\rm G}$. The conceptual diagram shows
gravitons with spin $S = \pm 2$~$(\pm 4 \times \hbar/2)$ and G$^+$ or G$^-$
orientation. It is unclear
whether such a spin would have any influence on the interaction process.
(Modified from Figure~1 of Paper~1)
\label{fig:grav_inter}}
\end{figure}

Spatially isolated particles at rest in an inertial system will be considered
first. The sum of the absorption and emission rates is set equal to the
intrinsic de Broglie frequency of the particle
\citep[cf. Schr\"odinger's Zitterbewegung; ][]
{Bro23,Sch30,Sch31,Hua52,Hes90,Pen06}.
Since the absorption and emission rates must be equal in
Eq.\,(\ref{eq.G_absorption}), this gives an emission coefficient of
%
%% Eq. 16
\begin{equation}
\eta_{\rm G} = \kappa_{\rm G}\,\rho_{\rm G} = \frac{1}{2}\frac{c^2_0}{h} =
6.782 \times 10^{49}\s^{-1}\kg^{-1} ~,% G\_eta
\label{eq.G_eta}
\end{equation}
i.e. half the intrinsic de Broglie frequency, since two virtual gravitons
are involved in each absorption/emission process
(cf. Fig.~\ref{fig:grav_inter}). The absorption coefficient is constant,
because both $\rho_{\rm G}$ and $\eta_{\rm G}$ are constant.
For an electron, for instance, with a mass of \\
$m_{\rm e} = 9.109\,383\,56(11) \times 10^{-31}\kg$, the virtual graviton
production rate equals its de Broglie frequency
$\nu^{\rm B}_{\rm G,e} = m_{\rm e}\,c_0^2/h = 1.235\,... \times 10^{20}\Hz$.

The energy absorption rate of an atomic particle with mass $M$ is
%
%% Eq. 17
\begin{equation}
\frac{\Del N_M}{\Del t}\,T^{\rm ab}_{\rm G} =
\kappa_{\rm G}\,\rho_{\rm G}\,M\,T_{\rm G}~.% G\_en\_abs
\label{eq.G_en_abs}
\end{equation}
Larger masses are thought of as conglomeration of atomic particle.

The emission energy, in turn, is assumed to be reduced to
%
%% Eq. 18
\begin{equation}
T_{\rm G}^{\rm em} = (1 - Y)\,T_{\rm G} % {eq.G\_reduction}
\label{eq.G_reduction}
\end{equation}
per graviton, where $Y$ ($0 < Y \ll 1$) is defined as the reduction
parameter. This leads to an energy emission rate of
%
%% Eq. 19
\begin{equation}
\frac{\Del N_M}{\Del t} \,T_{\rm G}^{\rm em} =
- \eta_{\rm G}\,M\,(1 - Y)\,T_{\rm G} ~ .% {eq.G\_emit}
\label{eq.G_emit}
\end{equation}
Without such an assumption the attractive gravitational force could not be
emulated, even with some kind of shadow effect as in Fatio's concept
\citep[cf.][]{Bop29}.
%
%% ONE-COLUMN FIGURE
% Figure 2
The reduction parameter~$Y$ and its relation to the attraction is
discussed below. If the energy-mass conservation \citep{Ein05c}
is applied, its consequence is that the mass of matter
increases with time at the expense of the background energy of the graviton
distribution.

A spherically symmetric emission of the liberated gravitons had been assumed
in Paper~1. Further studies summarized in Sects.\,\ref{ss.pot},
\ref{ss.galaxies} and \ref{ss.redshift} indicated that an anti-parallel
emission with respect to the incoming graviton has to be assumed in order to
avoid conflicts with energy and momentum conservation principles in closed
systems. This important assumption can best be explained by referring to
Fig.~\ref{fig:grav_inter}. The interaction is based on the combination of a
virtual graviton with momentum~$\vec{p}_{\rm G}$$^*$ and an incoming graviton
with $\vec{p}_{\rm G}$ followed by the liberation of another virtual graviton
in the opposite direction supplied with the excess energy~$T_{\rm G}^-$.
Regardless of the processes operating in the immediate environment of a
massive body, it must attract the mass of the combined real and virtual
gravitons, which will be at rest in the reference frame of the body.
The excess energy $T^-_{\rm G}$ is, therefore, reduced and so will be the
liberation energy, as assumed in Eq.\,(\ref{eq.G_reduction}). The emission
in Eq.\,(\ref{eq.G_emit}) will give rise to a flux of gravitons with reduced
energies in the environment of a body with mass~$M$. Its spatial density is
%
%% Eq. 20
\begin{equation}
\rho_M(r) = \frac{\Del N_M}{\Del V_r} =
\frac{\Del N_M}{\Del t}\,\frac{1}{4\,\pi\,r^2\,c_0} =
\eta_{\rm G}\,\frac{M}{4\,\pi\,r^2\,c_0} ~ ,% G\_Distr
\label{eq.G_Distr}
\end{equation}
where the volume increase is
%
%% Eq. 21
\begin{equation}
\Del V_r = 4\,\pi\,r^2\,c_0\,\Del t~. % volume
\label{eq.volume}
\end{equation}
The radial emission is part of the background in Eq.\,(\ref{eq.G_density}),
which has a larger number
density $\rho_{\rm G}$ than $\rho_M(r)$ at most distances, $r$, of interest.
Note that the emission of the gravitons from $M$ does not change the number
density or the total number of gravitons. For a
certain $r_M$, defined as the mass radius of $M$, it has to be
%
%% Eq. 22
\begin{equation}
\rho_{\rm G} =
\left [\frac{\Del N_M}
{\Del V_r} \right ]_{r_M}
= \frac{\eta_{\rm G}}{c_0}\,\frac{M}{4\,\pi\,r^2_M}~ ,% G\_approx
\label{eq.G_approx}
\end{equation}
because all gravitons of the background that come so close interact with the
mass $M$ in some way.
The same arguments apply to a mass $m \ne M$ and, in
particular, to the electron mass, $m_{\rm e}$. Therefore
%
%% Eq. 23
\begin{equation}
\sigma_{\rm G}
= \frac{m}{4\,\pi\,r^2_m}
= \frac{M}{4\,\pi\,r^2_M}
= \frac{m_{\rm e}}{4\,\pi\,r^2_{\rm G,e}}% ~G\_surface
\label{eq.G_surface}
\end{equation}
will be independent of the mass as long as the density of the background
distribution is constant.
The quantity $\sigma_{\rm G}$ is a kind of surface mass density. The equation
shows that $\sigma_{\rm G}$ is determined by the electron mass radius,
$r_{\rm G,e}$, for which estimates will be provided in Sects~\ref{ss.Pioneer}
and  \ref{ss.Earth_Sun}.
From Eqs.\,(\ref{eq.G_eta}),
(\ref{eq.G_approx}), and (\ref{eq.G_surface}), it follows that
%
%% Eq. 24
\begin{equation}
\kappa_{\rm G} \, \sigma_{\rm G} = c_0 ~ .% kappa\_sigma
\label{eq.kappa_sigma}
\end{equation}
The flux of modified gravitons from $M$ will
interact with a particle of
mass $m$ and vice versa. The interaction rate in the static case
can be found from Eqs.\,(\ref{eq.G_absorption}) and (\ref{eq.G_Distr}):
%
%% Eq. 25
\begin{eqnarray}
\frac{\Del N_{M,m}(r)}{\Del t} =
\kappa_{\rm G}\,m\,\frac{\Del N_M}{\Del V_r} =
\frac{\kappa_{\rm G}\,\eta_{\rm G}}{c_0}\,\frac{M\,m}{4\,\pi\,r^2} =
\nonumber \\
\frac{\kappa_{\rm G}\,c_0}{2\,h}\,\frac{m\,M}{4\,\pi\,r^2} =
\kappa_{\rm G}\,M\,\frac{\Del N_m}{\Del V_r} =
\frac{\Del N_{m,M}(r)}{\Del t} ~.% G\_interaction
\label{eq.G_interaction}
\end{eqnarray}

A calculation with anti-parallel emissions of the
secondary gravitons shows that an interaction of a graviton with reduced
momentum $\vec{p}_{\rm G}$$^-$ provides $-2\,\vec{p}_{\rm G}\,Y$ together with
its unmodified counterpart from the opposite sides. The resulting imbalance
will be
%
%% Eq. 26
\begin{eqnarray}
\frac{\Del \vec{P}_{M,m}(r)}{\Del t} =
- 2\,\vec{p}_{\rm G}\,Y\,\frac{\Del N_{M,m}(r)}{\Del t} =
\nonumber \\
- \vec{p}_{\rm G}\,Y\,\kappa_{\rm G}\,\frac{c_0}{h}\,\frac{M\,m}{4\,\pi\,r^2}~,
% G\_imbalance
\label{eq.G_imbalance}
\end{eqnarray}
if the quadratic terms in $Y$ can be neglected for very small $Y$ scenarios.

The imbalance will cause an attractive force that is
responsible for the gravitational pull between
bodies with masses $M$ and $m$. By comparing the force expression in
Eq.\,(\ref{eq.G_imbalance})
with Newton's law in Eq.\,(\ref{eq.Newton}), a relation between $p_{\rm G}$,
$Y$, $\kappa_{\rm G}$ and $G_{\rm N}$ can be established through the constant
$G_{\rm G}$:
%
%% Eq. 27
\begin{eqnarray}
G_{\rm G} = p_{\rm G}\,Y\,\kappa_{\rm G} = 4\,\pi\,G_{\rm N}\,\frac{h}{c_0} =
\nonumber \\
1.853\,... \times 10^{-51}\m^4\s^{-2}~.% G\_G
\label{eq.G_G}
\end{eqnarray}

\begin{figure}
\begin{center}
\includegraphics[width=\columnwidth]{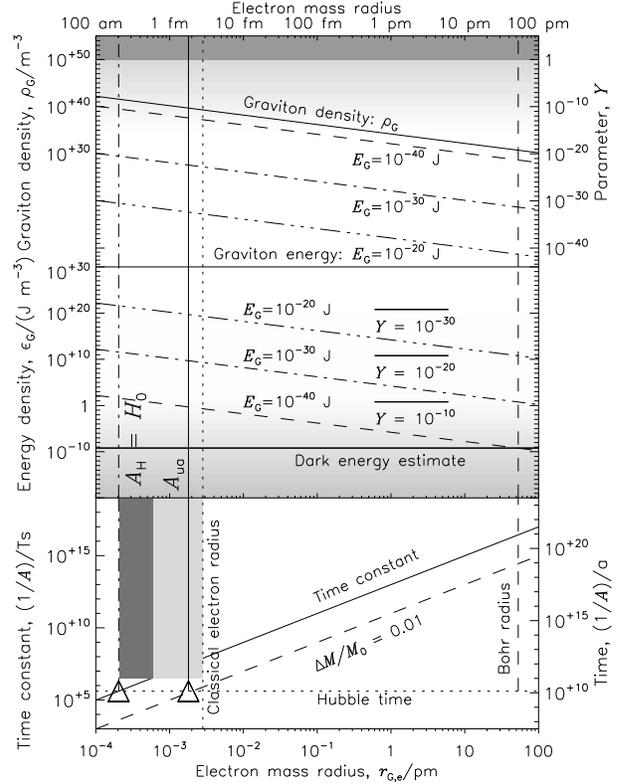}
\end{center}
\caption{Energies of $E_{\rm G} = (10^{-40}~{\rm to}~10^{-20})\J$ are assumed
for the gravitons, as indicated in the upper and middle panels by different
line styles. In the upper panel the spatial number density of gravitons and
the corresponding reduction parameter~$Y$ of
Eq.\,(\ref{eq.G_reduction}) are plotted as functions of the electron mass
radius~$r_{\rm G,e}$. The range $Y \ge 1$ (dark shading) is obviously
completely excluded by the model. Even values greater than $\approx 10^{-15}$
are not realistic (light shaded region),
cf. paragraph following Eq.\,(\ref{eq.G_G}).
The cosmic dark energy estimate
$(3.9 \pm 0.4)\GeV\m^{-3} = (6.2 \pm 0.6) \times 10^{-10}\J\m^{-3}$
\citep[see][]{BecMac}
is marked in the second panel.
It is well below the acceptable range (unshaded regions in the
middle panel).
If, however, only the $Y$ portion is taken into account in the dark energy
estimate, the total energy density could be many orders of magnitude larger as
shown for $Y$ from $10^{-10}$ to $10^{-30}$ by short horizontal bars.
In the lower panel, the mass accretion time constant and the time required
for a relative mass increase of 1~\% are shown (on the right side in units of
years). Indicated are also the Hubble time, $1/H_0$, as well as the lower
limit of the electron mass radius (left triangle and dark shaded area)
estimated from the Pioneer anomaly. The light shaded area takes smaller
Pioneer anomalies into account, see Sect.\,\ref{ss.Pioneer} .
It is shown up to the
vertical dotted line for the classical electron radius of 2.82\,fm. The right
triangle and the vertical solid line show the result in
Sect.\,\ref{ss.Earth_Sun}
based on the observed secular increase of the Sun-Earth distance
\citep{KraBru}. (Modified Figure~2 of Paper~1)
\label{fig:G_en_Y_H}}
\end{figure}

It can be seen that $Y$ does not depend on the mass of a body.
Since Eq.\,(\ref{eq.G_reduction}) allows stable processes over cosmological
time scales only, if $Y$ is very small, we assume in Fig.~\ref{fig:G_en_Y_H}
that $Y < 10^{-15}$.

Note that the mass of a body and thus its intrinsic de Broglie frequency are
not strictly constant in time, although the effect is only relevant for
cosmological time scales, see lower panel of Fig.~\ref{fig:G_en_Y_H}.
In addition, multiple interactions will occur within large mass
conglomerations (see Sects.\,\ref{ss.perihel} to \ref{ss.galaxies}), and can
lead to deviations from Eqs.\,(\ref{eq.Newton}).

The graviton energy density remote from any masses will be
%
%% Eq. 28
\begin{equation}
\epsilon_{\rm G} = T_{\rm G}\,\rho_{\rm G} =
\frac{2\,\pi\,G_{\rm N}}{Y}\,\sigma^2_{\rm G} ~% G\_en\_dens
\label{eq.G_en_dens}
\end{equation}
where the last term is obtained from Eq.\,(\ref{eq.G_G}) with the help of
Eqs.\,(\ref{eq.G_eta}) and (\ref{eq.G_approx}) to
(\ref{eq.kappa_sigma}).

What will be the consequences of the mass accretion required by the modified
model? With Eqs.\,(\ref{eq.G_en_abs}), (\ref{eq.G_emit}) and (\ref{eq.G_G}),
it follows that the relative mass accretion rate of a particle with mass $M$
will be

%
%% Eq. 29
\begin{equation}
A = \frac{1}{M}\,\frac{\Del M}{\Del t} =
\frac{2\,\pi\,G_{\rm N}}{c_0}\,\sigma_{\rm G} =
\frac{2\,\pi\,G_{\rm N}}{c_0}\,\frac{m_{\rm e}}{4\,\pi\,r^2_{\rm G,e}}~ ,
% G\_mass
\label{eq.G_mass}
\end{equation}
which implies an exponential growth according to
%
%% Eq. 30
\begin{equation}
M(t) = M_0\exp[A\,(t - t_0)] \approx M_0\,(1 + A\,\Del t) ~ ,% G\_growth
\label{eq.G_growth}
\end{equation}
where $M_0 = M(t_0)$ is the initial value at $t_0$ and
the linear approximation is valid for small $A\,(t - t_0) = A\,\Del t$.
The accretion rate is
%
%% Eq. 31
\begin{equation}
A = \frac{1.014 \times 10^{-49}}{r^2_{\rm G,e}}\m^2\s^{-1}~,% G\_accretion
\label{eq.G_accretion}
\end{equation}
if the expression is evaluated in terms of recent parameters.
%
%% ONE-COLUMN FIGURE
% Figure 3

With these assumptions, the gravitational quantities are displayed in
Fig.~\ref{fig:G_en_Y_H} in a wide parameter range (although the limits are set rather
arbitrarily). The lower panel displays the time constant of the mass
accretion. It indicates that a significant mass increase would be expected
within the standard age of the Universe of the order of $1/H_0$ (with
a Hubble constant of $H_0 \approx 2.43 \times 10^{-18}~\s^{-1}$) only
for very small $r_{\rm G,e}$.
\citet{FahHey} have suggested that a decay of the
vacuum energy density creates mass in an expanding Universe, and
\citet{Fah07} found a mass creation rate in accordance
with Eq.~(\ref{eq.G_growth}).

The relative uncertainty of the present knowledge of the Rydberg constant
%
%% Eq. 32
\begin{equation}
R_\infty = \frac{\alpha^2\,m_{\rm e}\,c_0}{2\,h} = 10\,973\,731.568\,508\m^{-1}
%Rydberg
\label{eq.Rydberg}
\end{equation}
is $u_{\rm r} \approx 5.9 \times 10^{-12}$,
where
%
%% Eq. 33
\begin{equation}
\alpha = \frac{e^2}{2\,\varepsilon_0\,c_0\,h} =
7.297\,352\,5664(17) \times 10^{-3}% ~eq.D\_alpha
\label{eq.D_alpha}
\end{equation}
is Sommerfeld's fine-structure constant.
Since spectroscopic observations of the distant Universe with redshifts up
to $z \le 0.5$ are compatible with modern data, it appears to be reasonable to
set $(1 + u_{\rm r})\,M_0 \ge M(t) > M_0$ at least for
$(t - t_0) \le 1.6 \times 10^{17}$~s. Any variation of $R_\infty$, caused by
the linear dependence upon the electron mass, which has also been considered by
\citet{Fah95},
would then be below the detection limit for state-of-the-art methods.

From the emission rate and the lifetime of virtual gravitons in
Eqs.\,(\ref{eq.G_absorption}) and (\ref{eq.G_lifetime}) an estimate of their
total number and energy at any time can thus be obtained for a body with
mass~$M$ as
%
%% Eq. 34
\begin{equation}
N^{\rm tot}_{\rm G} =
\Del t_{\rm G}\,\frac{M\,c^2_0}{h}%~ G\_tot\_num
\label{eq.G_tot_num}
\end{equation}
and
%
%% Eq. 35
\begin{equation}
T^{\rm tot}_{\rm G} = N^{\rm tot}_{\rm G}\,T^*_{\rm G}
\approx M\,c^2_0 ~ ,%~ G\_tot\_en
\label{eq.G_tot_en}
\end{equation}
i.e. the mass of a particle would reside within the virtual gravitons.

%%%%%%%%%%%%%%%%%%%%%%%%%%%%%%%%%%%%%%%%%%%%%%%%%%%%%%%%%%%%%%%%%%%%%%%%%%%%%%
%%%%%%%%%%%%%%%%%%%%%%%%%%%%%%%%%%%%%%%%%%%%%%%%%%%%%%%%%%%%%%%%%%%%%%%%%%%%%%

\subsection{Coulomb's law}
\label{ss.Coulomb}
%% Subsect. 2.5

%%%%%%%%%%%%%%%%%%%%%%%%%%%%%%%%%%%%%%%%%%%%%%%%%%%%%%%%%%%%%%%%%%%%%%%%%%%%%%

\subsubsection{Electric fields and charged particles}
\label{sss.E_interact}
%% Sect. 2.5.1

Coulomb's law in Eq.\,(\ref{eq.Coulomb}) gives the
attractive or repulsive electrostatic force between two charged particles
at rest in an inertial system. Together with the electric field in
Eq.\,(\ref{eq.E_field}) it can be written as
%
% Eq. 36
\begin{equation}
\vec{K}_{\rm E}(r) = \vec{E}_Q(r)\,q ~.% D\_el
\label{eq.D_el}
\end{equation}
The electric potential, $\Ph_Q(r)$, of a charge, $Q$, located at $r = 0$ is
%
% Eq. 37
\begin{equation}
\Ph_Q(r) =
\frac{Q}{4\,\pi\,r\,\varepsilon_0} %~ D\_pot
\label{eq.D_pot}
\end{equation}
for $r > 0$. The corresponding electric field can thus be written as
$\vec E_Q(r) = - \nabla \Ph_Q(r)$.

%%%%%%%%%%%%%%%%%%%%%%%%%%%%%%%%%%%%%%%%%%%%%%%%%%%%%%%%%%%%%%%%%%%%%%%%%%%%%%

\subsubsection{Dipole interactions}
\label{sss.D_interact}
%% Sect. 2.5.2

Note that the dipoles in the background distribution,
cf. Eq.\,(\ref{eq.D_density}),
have no mean interaction energy, even in the classical theory
\citep[see e.\,g.][]{Jac06}.
Whether this ``background dipole radiation'' and the ``graviton radiation''
are related to the dark matter (DM) and dark energy (DE) problems is of no
concern here, but could be an interesting speculation.
%
%% ONE-COLUMN FIGURE
% Figure 4
\begin{figure}[t]
\begin{center}
\includegraphics[width=\columnwidth]{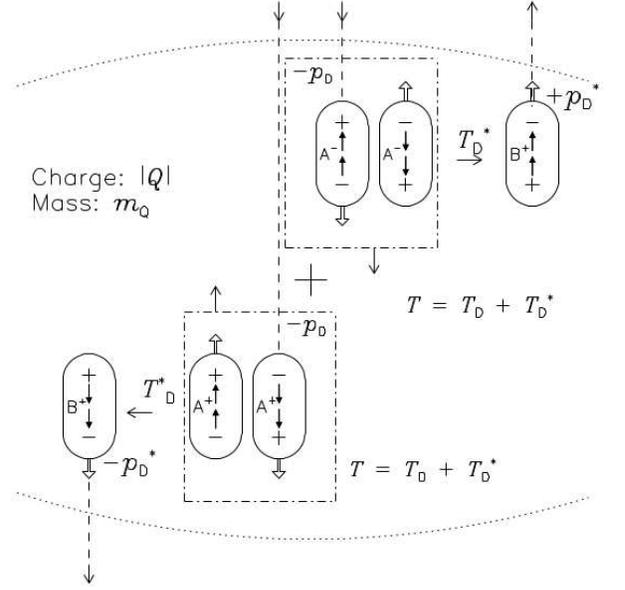}
\end{center}
\caption{Virtual dipoles emitted by a charge $+|Q|$ (with mass $m_Q$)
interact with ``real'' dipoles, A$^+$ and A$^-$ arriving with a
momentum~$-\,p_{\rm D}$, each.
On the left, a dipole~A$^+$ combines in the lower
dashed-dotted box with virtual dipole~A$^+$ in its destruction phase
and
liberates dipole~B$^+$. \emph{No momentum
will be transferred to the central charge} with $p_{\rm D}  = p^*_{\rm D} $.
The other type of
interaction\,--\,called direct interaction,
in contrast to the indirect one on the left\,--\,also
requires two virtual dipoles, one of them combines in its creation phase
with
dipole~A$^-$ (in the upper box with dashed-dotted boundaries),
the other one is liberated by the excess energy of the annihilation.
\emph{The central charge received a momentum of}
$ - (p_{\rm D}  + p^*_{\rm D} )$. No spin reversal has been assumed
in both cases.
\label{fig:D_interaction}}
\end{figure}
A charge, $Q$, absorbs and emits dipoles at a rate
%
% Eq. 38
\begin{equation}
\frac{\Del N_Q}{\Del t}
= \kappa_{\rm D}\,\rho_{\rm D}\,|Q|
= \eta_{\rm D}\,|Q|
~ ,% ~D\_Emission
\label{eq.D_Emission}
\end{equation}
where $\eta_{\rm D}$ and $\kappa_{\rm D}$ are the
corresponding (dipole) emission and absorption
coefficients.

From energy conservation it follows that absorption and emission rates
of dipoles in Eq.\,(\ref{eq.D_Emission}) of a body with charge~$Q$
must be equal.
The momentum conservation can, in general, be fulfilled by isotropic
absorption and emission processes.

The interaction processes assumed between a positively charged body and
dipoles is sketched in Fig.~\ref{fig:D_interaction}.
A mass~$m_Q$ of the charge~$Q$ has explicitly been mentioned,
because the massless dipole charges are not assumed to absorb and emit
any dipoles themselves. The conservation of momentum could hardly be
fulfilled in such a process. In Sect.\,\ref{ss.redshift} we postulate,
however, that gravitons interact with dipoles and thereby control their
momentum and speed, subject to the condition that $p_{\rm G} \ll p_{\rm D}$.

The assumptions
as outlined will lead to a distribution of the emitted dipoles in the rest
frame of an isolated charge, $Q$, with a spatial density of
%
% Eq. 39
\begin{equation}
\rho_Q(r) = \frac{\Del N_Q}{\Del V_r} =
\frac{1}{4\,\pi\,r^2\,c_0}\,\frac{\Del N_Q}{\Del t} =
\eta_{\rm D}\,\frac{|Q|}{4\,\pi\,r^2\,c_0} ~ ,% D\_Distribution
\label{eq.D_Distribution}
\end{equation}
where $\Del V_r$ is given in Eq.\,(\ref{eq.volume}).
The radial emission is part of the background~$\rho_{\rm D}$,
which has a larger number density
than $\rho_D(r)$
at most distances, $r$, of interest.
Note that the emission of the
dipoles from $Q$ does not change the number density,
$\rho_{\rm D}$, in the environment of the charge,
but reverses the orientation of \emph{half} of the dipoles affected.

The total number of dipoles will, of course, not be changed either.
For a certain $r_Q$, defined
as the charge radius of $Q$, it has to be
%
% Eq. 40
\begin{equation}
\rho_{\rm D} =
\left [\frac{\Del N_Q}
{\Del V_r} \right ]_{r_Q}
= \frac{\eta_{\rm D}}{c_0}\,\frac{|Q|}{4\,\pi\,r^2_Q}
~ ,% D\_approximation
\label{eq.D_approximation}
\end{equation}
because all dipoles of the background that come so close interact
with the charge $Q$ in some way.
The same arguments apply to a charge $q \ne Q$. Since $\rho_{\rm D}$
cannot depend on either $q$ or $Q$, the quantity
%
% Eq. 41
\begin{equation}
\sigma_{\rm Q}
= \frac{|Q|}{4\,\pi\,r^2_Q}
= \frac{|q|}{4\,\pi\,r^2_q}
= \frac{|e|}{4\,\pi\,r^2_e} % ~ D\_surface
\label{eq.D_surface}
\end{equation}
must be independent of the charge, and can be considered as a kind of
surface charge density,
cf. ,,Fl\"achenladung'' of an electron defined by \citet{Abr02},
that is the same for all charged particles.
The equation shows that $\sigma_{\rm Q}$ is determined by the electron
charge radius, $r_{\rm e}$.

At this stage, this is a formal description
awaiting further quantum electrodynamic studies in the near-field region
of charges. It might, however, be instructive to provide a speculation
for the dipole emission rate~$\Del N_Q /\Del t$ of a charge~$Q$.
The physical constants~$\alpha$, $c_0$, $h$, $\epsilon_0$ and $G_{\rm N}$
can be combined to give a dipole emission coefficient
%
% Eq. 42
\begin{equation}
\eta_{\rm D} = \frac{1}{2}\frac{\alpha^2\,c_0^2}{h\,\sqrt{\epsilon_0\,G_{\rm N}}} =
 1.486 \times 10^{56}\s^{-1}\C^{-1}
\label{eq.eta_D}
\end{equation}
as half the virtual dipole production rate and thus for a charge~$|e|$ a rate
%
% Eq. 43
\begin{eqnarray}
\frac{\Del N_e}{\Del t} = \eta_{\rm D}\,|e| =  2.380 \times 10^{37}\s^{-1}~.
\label{eq.N_t_D}
\end{eqnarray}
Note that dipole emission rate, in contrast to the assumptions in Fig.~5
of Paper~2, is fixed for a certain charge and does not
depend on the particle mass.

From Eqs.\,(\ref{eq.D_Emission}), (\ref{eq.D_approximation}), and
(\ref{eq.D_surface}) we get

% Eq. 44
\begin{equation}
\kappa_{\rm D}\,\sigma_{\rm Q} = c_0 %~ . D\_kappa
\label{eq.D_kappa}
\end{equation}
During a \emph{direct} interaction, the dipole~A$^-$
(in Fig.~\ref{fig:D_interaction}
on the right side) combines together
with an identical virtual dipole with an opposite velocity
vector. This postulate is motivated by the fact that it provides the
easiest way to eliminate the charges and yield
$P = -\,p_{\rm D}  + p^*_{\rm D}  = 0$ (where $p^*_{\rm D} $ is the momentum
of the virtual dipole) as well as $S = 0$.
The momentum balance is neutral and the
excess energy, $T_{\rm D} $, is used to liberate a second virtual dipole~B$^+$,
which has the required orientation.
The charge had emitted two virtual dipoles
with a momentum of $+\,p^*_{\rm D} $, each, and a total momentum of
$- (p_{\rm D}  + p^*_{\rm D} )$ was transferred to $|Q|$.
The process can be described as a reflection of a dipole
together with a reversal of the dipole momentum.
The number of these direct interactions will be denoted by
$\Del \hat{N}_Q$.
The dipole of type~A$^+$ (on the left side) can
exchange its momentum in an \emph{indirect} interaction
only on the far side of the charge with an identical virtual dipole
during its absorption (or destruction) phase (cf.
Fig.~\ref{fig:Creation}).
The excess energy of $T_{\rm D} $
is supplied to liberate a second virtual dipole~B$^+$.
The momentum transfer to the charge $+\,|Q|$ is zero.
This process just corresponds to a
double charge exchange. Designating the number of interactions of the
indirect type with
$\Del \tilde{N}_Q$, it is
%
% Eq. 45
\begin{equation}
\frac{\Del N_Q}{\Del t} =
\frac{\Del \hat{N}_Q +
\Del \tilde{N}_Q}{\Del t}% ~ ,Distrib2
\label{eq.Distrib2}
\end{equation}
with $\Del \tilde{N}_Q = \Del \hat{N}_Q =
\Del N_Q/2$. Unless direct and indirect interactions
are explicitly specified, both types are meant by the term ``interaction''.

The virtual dipole emission rate in Fig.~\ref{fig:D_interaction} has to be
%
% Eq. 46
\begin{equation}
\frac{\Del N^*_Q}{\Del t} =
2\,\frac{\Del N_Q}{\Del t} ~ ,%virtualrate
\label{eq.virtualrate}
\end{equation}
i.e., the virtual dipole emission rate equals the sum of the real
absorption and emission rates.
The interaction model described results in a mean momentum
transfer per interaction of $p_{\rm D}$
\emph{without involving a macroscopic electrostatic field}.

A quantitative evaluation gives
the force acting on a test particle with
charge, $q$, at a distance, $r$, from another particle with charge
$Q$. This results from the absorption of dipoles not only from the background,
but also from the distribution emitted from $Q$ according to
Eq.\,(\ref{eq.D_Distribution}) under the assumption of a \emph{constant}
absorption coefficient, $\kappa_{\rm D}$ in Eq.\,(\ref{eq.D_Emission}).
The rate of interchanges between these charges then is
%
% Eq. 47
\begin{eqnarray}
\frac{\Del N_{Q,q}(r)}{\Del t} =
\kappa_{\rm D}\,|q|\,\frac{\Del N_Q}{\Del V_r} =
\frac{\kappa_{\rm D}\,\eta_{\rm D}}{c_0}\,\frac{|Q|\,|q|}{4\,\pi\,r^2} =
\nonumber \\
\kappa_{\rm D}\,|Q|\,\frac{\Del N_q}{\Del V_r} =
\frac{\Del N_{q,Q}(r)}{\Del t} ~ ,% D\_reciprocal
\label{eq.D_reciprocal}
\end{eqnarray}
which confirms the reciprocal relationship between $q$ and $Q$.
The equation, also very similar to Eq.\,(\ref{eq.G_interaction}),
does not contain an explicit value for~$\eta_{\rm D}$.
It is important to realize that all interchange events between pairs of
charged particles are either direct or indirect depending on their
polarities and transfer a momentum of $2\,p_{\rm D}$ or zero.

The external electrostatic potential of a spherically symmetric body~C with
charge $Q$ is given in Eq.\,(\ref{eq.D_pot}).
Since the electrostatic forces between the charged particles~C and D
are typically many orders of magnitude larger than the gravitational forces,
we only take the electrostatic effects into account in this section and
neglect the gravitational interaction.

\begin{figure}[!t]
% Figure 5
\centering
\includegraphics[width=7.6cm]{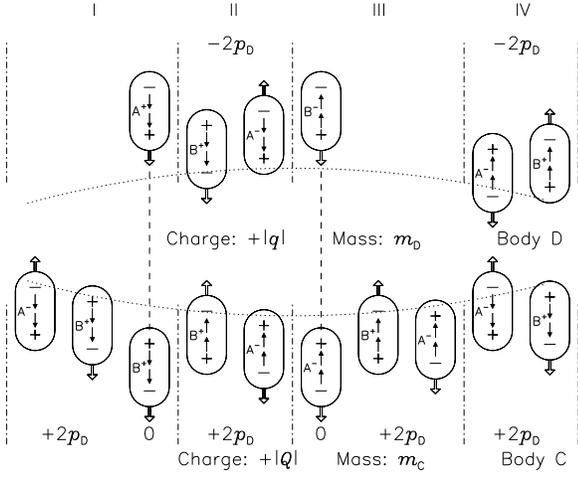}
\caption{\label{fig:plus} The body~C with charge~$Q > 0$ and mass~$m_{\rm C}$
is positioned in this configuration beneath body~D with charge~$+|q|$ and
mass~$m_{\rm D}$ leading to an
electrostatic repulsion of the bodies. This results from
the reversal of dipoles by the charge~$Q$ followed by {\it direct}
interactions with the charge~$+|q|$ as defined on the right-hand side of
Fig.~\ref{fig:D_interaction}.
\emph{Two reversals are schematically indicated in columns~I and III.} The
dipoles arriving in columns~II and IV from below have the same polarity as if
they would be part of the background distribution and do not contribute
to the momentum transfer, because of a compensation by
dipoles arriving from above. The net momentum transfer caused by the two
interacting reversed dipoles thus is $4\,\vec{p}_{\rm D}$,
i.e. $2\,\vec{p}_{\rm D}$ per dipole. (Modified Figure~3 of Paper~2)}
\end{figure}

\begin{figure}[!t]
% Figure 6
\centering
\includegraphics[width=7.6cm]{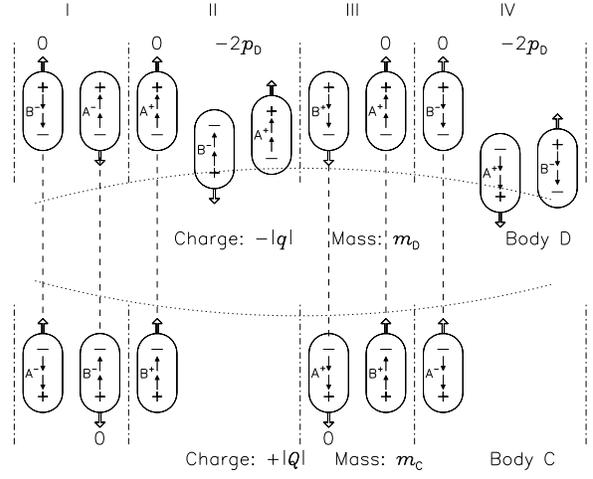}
\caption{\label{fig:minus} The body~C is again positioned
beneath body~D. The charge of D is now~$-|q|$, however, leading to an
electrostatic attraction of the bodies. The attraction results from
the reversal of dipoles by the charge~$Q > 0$ followed by {\it indirect}
interactions with charge~$-|q|$ as defined on the left-hand side of
Fig.~\ref{fig:D_interaction}.
\emph{Two events without momentum transfer are schematically indicated
in columns ~II and IV.} The
dipoles arriving in columns~I and III from below have the same polarity as if
they would be part of the background distribution. The same is true for all
dipoles arriving from above. The net momentum transfer caused by the two
reversed dipoles thus is $-4\,\vec{p}_{\rm D}$,
i.e. $-2\,\vec{p}_{\rm D}$ per
dipole. (Modified Figure~4 of Paper~2)}
\end{figure}

In order to have a well-defined configuration for our discussion,
we will assume that body~C with mass $m_{\rm C}$ has a positive
charge~$Q > 0$ and is positioned at a distance~$r$
beneath body~D (mass $m_{\rm D}$) with
either a charge $+|q|$ in Fig.~\ref{fig:plus} or
$-|q|$ in Fig.~\ref{fig:minus}. Only the processes near the body~D are shown
in detail.


The interaction rates of dipoles with bodies~C and D in
Eq.\,(\ref{eq.D_reciprocal})
(the same for both bodies even if $|Q| \ne |q|$)
and the momentum transfers indicated in
Figs.~\ref{fig:plus} and \ref{fig:minus}, respectively,
lead to a norm of the momentum change rate
for bodies C and D of
%
%% Eq. 48
\begin{equation}
\left|\frac{\Del\vec{P}_{\rm Q,q}(r)}{\Del t}\right| =
2\,p_{\rm D}\,\frac{\Del N_{Q,q}(r)}{\Del t} = %\nonumber \\
2\,p_{\rm D}\,\frac{\kappa_{\rm D}\,\eta_{\rm D}}{c_0}\,\frac{Q\,|q|}
{4\,\pi\,r^2}  ~ .% imbalance
\label{eq.imbalance}
\end{equation}
Together with
%
%% Eq. 49
\begin{equation}
p_{\rm D}\,\frac{\kappa_{\rm D}\,\eta_{\rm D}}{c_0} =
\frac{T_{\rm D}\,\kappa_{\rm D}\,\eta_{\rm D}}{c^2_0} =
\frac{1}{2\,\varepsilon_0}% ~ D\_relation
\label{eq.D_relation}
\end{equation}
this leads, depending on the signs of the charges~$Q$ and $q$, to a repulsive
or an attractive electrostatic force between C and D
in accordance with Coulomb's law in Eq.\,(\ref{eq.Coulomb}).

Important questions are related to the energy~$T_{\rm D}$ and
momentum~$\vec{p}_{\rm D}$ of the
dipoles and, even more, to their energy density
in space.
Eqs.\,(\ref{eq.D_energy}), (\ref{eq.D_approximation}),
(\ref{eq.D_surface}), and (\ref{eq.D_kappa}) together with
Eq.\,(\ref{eq.D_relation}) allow the energy density to be expressed by
%
% Eq. 50
\begin{equation}
\epsilon_{\rm D} = T_{\rm D}\,\rho_{\rm D} =
\frac{\sigma^2_{\rm D}}{2\,\varepsilon_0} ~ .% eq.energydensity
\label{eq.energydensity}
\end{equation}
This quantity is independent of the dipole energy.
It takes into account all
dipoles (whether their distribution is chaotic or not). Should the energy
density vary in space and\,/\,or time, the surface charge density,
$\sigma_{\rm Q}$, must vary as well.

If we assume that the electron charge radius $r_{\rm Q}$ in
Eq.\,(\ref{eq.D_surface}) equals
the classical electron radius 2.82~fm,
very high energy densities of
$\epsilon_{\rm D}~=~1.45~\times~10^{29}\J\m^{-3}$, compared to the cosmic
dark energy estimate, follow from Eq.\,(\ref{eq.energydensity}).
The dipole density in Eq.\,(\ref{eq.D_approximation}) is also very high
with $\rho_{\rm D} =  7.95  \times 10^{56}\m^{-3}$ leading to a dipole
energy of $T_{\rm D} = 1.83 \times 10^{-28}\J$.
If, on the other hand, we identify the dipole distribution with DM
with an estimated energy density of $2.48 \times 10^{-10}\J\m^{-3}$ and
require that the dipole energy density corresponds to this value, then
rather unlikely values follow for
$r_{\rm Q} = 13.9\,\mu{\rm m}$,
$\rho_{\rm D} = 3.28 \times 10^{37}\m^{-3}$ and
$T_{\rm D} = 7.55 \times 10^{-48}\J$.

%%%%%%%%%%%%%%%%%%%%%%%%%%%%%%%%%%%%%%%%%%%%%%%%%%%%%%%%%%%%%%%%%%%%%%%%%%%%%%
%%%%%%%%%%%%%%%%%%%%%%%%%%%%%%%%%%%%%%%%%%%%%%%%%%%%%%%%%%%%%%%%%%%%%%%%%%%%%%
%%%%%%%%%%%%%%%%%%%%%%%%%%%%%%%%%%%%%%%%%%%%%%%%%%%%%%%%%%%%%%%%%%%%%%%%%%%%%%
%%%%%%%%%%%%%%%%%%%%%%%%%%%%%%%%%%%%%%%%%%%%%%%%%%%%%%%%%%%%%%%%%%%%%%%%%%%%%%

\section{Applications of impact models}
\label{s.impact}
%% Sect. 3

%%%%%%%%%%%%%%%%%%%%%%%%%%%%%%%%%%%%%%%%%%%%%%%%%%%%%%%%%%%%%%%%%%%%%%%%%%%%%%
%%%%%%%%%%%%%%%%%%%%%%%%%%%%%%%%%%%%%%%%%%%%%%%%%%%%%%%%%%%%%%%%%%%%%%%%%%%%%%

The detection of gravitons and dipoles with the expected properties would,
of course, be the best verification of the proposed models. Laking this,
indirect support can be found through the application of the models with a
view to describe physical processes successfully for specific situations.

\subsection{Potential energies}
\label{ss.pot}
%% Subsect. 3.1

%%%%%%%%%%%%%%%%%%%%%%%%%%%%%%%%%%%%%%%%%%%%%%%%%%%%%%%%%%%%%%%%%%%%%%%%%%%%%%

\subsubsection{Gravitational potential energy}
\label{sss.Gravity}
%% Subsect. 3.1.1

As mentioned in Sect.\,\ref{s.intro} the study of the potential energy
problem \citep{WilDwi15a} had been
motivated by the remark that the potential energy is rather
mysterious \citep{Car98}\footnote{In this context it is of interest
that \citet{Bri65} discussed this problem in relation to electrostatic
potential energy.}. It led to the identification of the
``source region'' of the potential energy for the special case of
a system with two masses~$M_{\rm E}$ and $M_{\rm M}$ subject to the condition
$M_{\rm E} \gg M_{\rm M}$. An attempt to generalize the study without this
condition required either violations of the energy conservation principle
as formulated by \citet{Lau20} for a closed system, or a reconsideration
of an assumption we made concerning the gravitational interaction
process in Paper~1. The change necessary to comply with the energy
conservation principle has been discussed in Sect.\,\ref{ss.Newton}.
A generalization of the potential energy concept for a system of two
spherically symmetric bodies~A
and B with masses~$m_{\rm A}$ and $m_{\rm B}$ without the above condition
could then be formulated \citep{WilDwi15b}.

We will again exclude any further energy contributions, such as rotational or
thermal energies, and make use of the fact that the external gravitational
potential of a spherically symmetric body of mass~$M$ and radius~$r$ in
Eq.\,(\ref{eq.G_potential}) is
that of a corresponding point mass at the centre.

The energy~$E_m$ and the momentum~$\vec{p}$ of a free particle with mass~$m$
moving with a velocity~$\vec{v}$ relative to an inertial reference system are
related by
%
%% Eq. 51
\begin{equation}
E^2_m - \vec{p}^{\,2}\,c^2_0 = m^2\,c^4_0 ~,
\label{eq.energy}
\end{equation}
where $\vec{p}$ is the momentum vector
%
%% Eq. 52
\begin{equation}
\vec{p} = \vec{v}\,\frac{E_m}{c^2_0}~.
\label{eq.momentum}
\end{equation}
\citep{Ein05b,Ein05c}.
For an entity in vacuum with no rest mass ($m = 0$), such as a
photon \citep[cf.][]{Ein05a,Lew26,Oku09}, the energy-momentum
relation in Eq.\,(\ref{eq.energy}) reduces to
%
%% Eq. 53
\begin{equation}
E_\nu = p_\nu\,c_0 ~.
\label{eq.photon}
\end{equation}

We now assume that two spherically symmetric bodies~A and B with masses~$m_{\rm A}$
and $m_{\rm B}$, respectively, are placed in space remote from other
gravitational centres at a distance of~$r + \Del r$ reckoned from the position
of~A. Initially both bodies are at rest with respect to an inertial reference
frame represented by the centre of gravity of both bodies. The total energy of
the system then is with Eq.\,(\ref{eq.energy}) for the rest energies and
with Eq.\,(\ref{eq.G_potential}) for the potential energy
%
%% Eq. 54
\begin{equation}
E_{\rm S} = (m_{\rm A} + m_{\rm B})\,c^2_0 -
G_{\rm N}\,\frac{m_{\rm A}\,m_{\rm B}}{r + \Del r} ~.
\label{eq.total}
\end{equation}
The evolution of the system during the approach of A and B from $r +\Del r$ to
$r$ can be described in classical mechanics.
According to Eq.\,(\ref{eq.imbalance}), the attractive force between the
bodies during the approach is approximately constant  for $r \gg \Del r > 0$,
resulting in accelerations of $b_{\rm A} = |K_{\rm G}(r)|/m_{\rm A}$ and
$b_{\rm B} = - |K_{\rm G}(r)|/m_{\rm B}$, respectively. Since the duration~$\Del t$ of
the free fall of both bodies is the same, the approach of A and B can be
formulated as
%
%% Eq. 55
\begin{eqnarray}
\Del r = s_{\rm A} - s_{\rm B} =
\frac{1}{2}\,(b_{\rm A} - b_{\rm B})\,(\Del t)^2 = \nonumber \\
\frac{1}{2}\,\left(\frac{1}{m_{\rm A}} +
\frac{1}{m_{\rm B}}\right)\,|K_{\rm G}(r)|\,(\Del t)^2 ~,% G\_acc
\label{eq.G_acc}
\end{eqnarray}
showing that $s_{\rm A}\,m_{\rm A} = - s_{\rm B}\,m_{\rm B}$, i.e, the centre
of gravity stays at rest. Multiplication of Eq.\,(\ref{eq.G_acc}) by~$|K_{\rm G}(r)|$
gives the corresponding kinetic energy equation
%
%% Eq. 56
\begin{eqnarray}
|K_{\rm G}(r)|\,\Del r =
\frac{1}{2}\,\left(\frac{K^2_{\rm G}(r)\,(\Del t)^2}{m_{\rm A}} +
\frac{K^2_{\rm G}(r)\,(\Del t)^2}{m_{\rm B}}\right) = \nonumber \\
\frac{1}{2}\,m_{\rm A}\,v_{\rm A}^2 +
\frac{1}{2}\,m_{\rm B}\,v_{\rm B}^2 = T_{\rm A} + T_{\rm B} ~.% G\_accel
\label{eq.G_accel}
\end{eqnarray}
The kinetic energies\footnote{Eqs.\,(\ref{eq.energy}) and (\ref{eq.momentum})
together with $E_0 = m\,c^2_0$ \citep{Ein05c} and
$\gamma = 1/\sqrt{1 - v^2/c^2_0}$ yield the relativistic kinetic energy of a
massive body: $T = E - E_0 = E_0\,(\gamma - 1)$. The evaluations for
$T_{\rm A}$ and $T_{\rm B}$ agree in very good approximation with
Eq.\,(\ref{eq.G_accel}) for small $v_{\rm A}$ and
$v_{\rm B}$.\label{footnote_ein}} $T_{\rm A}$ and $T_{\rm B}$
should, of course, be the difference of the potential
energy in Eq.\,(\ref{eq.total}) at distances of
$r$ and $r + \Del r$. We find indeed with Newton's law in Eq.\,(\ref{eq.Newton})
%
%% Eq. 57
\begin{eqnarray}
G_{\rm N}\,m_{\rm A}\,m_{\rm B}\left(\frac{1}{r} - \frac{1}{r + \Del r}\right)
\approx \nonumber \\
G_{\rm N}\,\frac{m_{\rm A}\,m_{\rm B}}{r^2}\,\Del r = |K_{\rm G}(r)|\,\Del r ~.
% pot\_diff
\label{eq.pot_diff}
\end{eqnarray}

We may now ask the question, whether the impact model can provide an answer to
the potential energy ``mystery'' in a closed system.
Since the model implies a secular increase of mass of all
bodies, it obviously violates a closed-system assumption. The
increase is, however, only
significant over cosmological time scales, and we can neglect its consequences
in this context. A free single body will, therefore, still be considered as a
closed system with constant mass. In a two-body system both
masses~$m_{\rm A}$ and $m_{\rm B}$ will be constant in such an approximation,
but now there are gravitons interacting with both masses.

The number of gravitons travelling at any instant of time from one mass to
the other can be calculated from the interaction
rate in Eq.\,(\ref{eq.G_interaction}) multiplied by the travel time~$r/c_0$:
%
%% Eq. 58
\begin{equation}
\Del N_{m_{\rm A},m_{\rm B}}(r) =
\frac{\kappa_{\rm G}}{8\,\pi\,h}\,\frac{m_{\rm A}\,m_{\rm B}}{r} ~.% eq.number
\label{eq.number}
\end{equation}
The same number is moving in the opposite direction.
The energy deficiency of the interacting gravitons with respect to the
corresponding background then is together with Eqs.\,(\ref{eq.G_reduction})
and (\ref{eq.G_G}) for each body
%
%% Eq. 59
\begin{eqnarray}
\Del E_{\rm G}(r) =
-  p_{\rm G}\,Y\,\kappa_{\rm G}\,\frac{c_0}
{8\,\pi\,h}\,\frac{m_{\rm A}\,m_{\rm B}}{r} =
\nonumber \\
- G_{\rm G}\,\frac{c_0}{8\,\pi\,h}\,\frac{m_{\rm A}\,m_{\rm B}}{r} =
- \frac{G_{\rm N}}{2}\,\frac{m_{\rm A}\,m_{\rm B}}{r} ~.% eq.deficiency
\label{eq.deficiency}
\end{eqnarray}
The last term shows\,--\,with reference to Eq.\,(\ref{eq.pot_diff})\,--\,that
the energy deficiency~$\Del E_{\rm G}$ equals {\em half} the potential energy
of body~A at a distance~$r$ from body~B and vice versa.

We now apply Eq.\,(\ref{eq.deficiency}) and calculate the difference
of the energy deficiencies for separations of $r + \Del r$ and $r$ for
interacting gravitons travelling in both directions and get
%
%% Eq. 60
\begin{eqnarray}
2\,\{\Del E_{\rm G}(r + \Del r) - \Del E_{\rm G}(r)\} = \nonumber \\
G_{\rm N}\,m_{\rm A}\,m_{\rm B}\,\left(\frac{1}
{r} - \frac{1}{r + \Del r}\right)~.% eq.diff\_def
\label{eq.diff_def}
\end{eqnarray}
Consequently, the difference of the potential energies between
$r + \Del r$ and $r$ in Eq.\,(\ref{eq.pot_diff}) is balanced
by the difference of the total energy deficiencies.

The physical processes involved can be described as follows:
\begin{enumerate}
\item The number of gravitons on their way for a separation of $r + \Del r$
is smaller than that for $r$, because the interaction rate
depends on $r^{-2}$ according to Eq.\,(\ref{eq.imbalance}), whereas the
travel time is proportional to $r$.
\item A decrease of~$r + \Del r$ to $r$ during the approach of~A and B
increases the number of gravitons with reduced energy.
\item The energies liberated by energy reductions are
available as potential energy and are converted into kinetic energies
of the bodies~A and B.
\item With Eqs.\,(\ref{eq.energy}) and (\ref{eq.momentum}) and the
approximations in Footnote~\ref{footnote_ein}, it follows that the sum of the
kinetic energies~$T_{\rm A}$ and $T_{\rm B}$, the masses~A and B plus
the total energy deficiencies of the interacting gravitons can indeed
be considered to be a closed system as defined by \citet{Lau20}.
\end{enumerate}

%%%%%%%%%%%%%%%%%%%%%%%%%%%%%%%%%%%%%%%%%%%%%%%%%%%%%%%%%%%%%%%%%%%%%%%%%%%%%%

\subsubsection{Electrostatic potential energy}
\label{sss.elec}
%% Subsect. 3.1.2

%\citet{WilDwi14c}%2

In this section we will discuss the electrostatic aspects of
the potential energy.

The energy density of an electric field~$\vec E$ outside of charges is given
by
%
% Eq. 61
\begin{equation}
w = \frac{\varepsilon_0}{2}\,\vec E^2% ~ D\_sta\_en
\label{eq.D_sta_en}
\end{equation}
\citep[cf., e.g.][]{Hun57,Jac06}. Applying Eq.\,(\ref{eq.D_sta_en})
to a plane plate capacitor with an area~$F$,
a plate separation~$b$
and charges $\pm\,|Q|$ on the plates, the energy stored in the field
of the capacitor turns out to be
%
% Eq. 62
\begin{equation}
W = \frac{\varepsilon_0}{2}\,\vec E^2\,F\,b =
\frac{\varepsilon_0}{2}\,\vec E^2\,V ~ .% D\_capacitor
\label{eq.D_capacitor}
\end{equation}
With a potential difference $\Del\Ph = |\vec E|\,b$
and a charge of $Q = \varepsilon_0\,|\vec E|\,F$
(incrementally increased to these values), the potential energy
of $Q$ at $\Del\Ph$ is
%
% Eq. 63
\begin{equation}
W = \frac{1}{2}\,Q\,\Del\Ph ~ .% D\_ca\_pot\_en
\label{eq.D_ca_pot_en}
\end{equation}
The question as to where the energy is actually stored,
\citet{Hun57}
answered by showing
that both concepts implied by
Eqs.\,(\ref{eq.D_capacitor}) and (\ref{eq.D_ca_pot_en}) are equivalent.

Can the impact model provide an answer for the electrostatic potential energy
in a closed system, where dipoles are interacting with two charged bodies?
The number of reversed dipoles travelling at any instant of time from a
charge~$Q > 0$ to $q$ in Figs.~\ref{fig:plus} and \ref{fig:minus} can be
calculated from the interaction
rate in Eq.\,(\ref{eq.D_reciprocal}) multiplied by a travel
time~$\Del t = r/c_0$:
%
%% Eq. 64
\begin{equation}
\Del N_{Q,q}(r) =
\frac{\kappa_{\rm D}\,\eta_{\rm D}}{c_0^2}\,\frac{Q\,|q|}{4\,\pi\,r}  ~.
% D\_number
\label{eq.D_number}
\end{equation}
The same number of dipoles is moving in the opposite direction from $q$ to
$Q$. From Eqs.\,(\ref{eq.D_energy}), (\ref{eq.D_relation}) and
(\ref{eq.D_number}), we can
determine the total energy of the reversed dipoles:
%
%% Eq. 65
\begin{equation}
\Del E_{Q,q}(r) = 2\,\Del N_{Q,q}(r)\,p_{\rm D}\,c_0 =
\frac{Q\,|q|}{4\,\pi\,\varepsilon_0\,r}  ~.% pot\_energy
\label{eq.pot_energy}
\end{equation}
It is equal to the absolute value of the electrostatic potential energy of
a charge~$q$ at the electric potential~$\Ph_Q(r)$
in Eq.\,(\ref{eq.D_pot}) of a charge~$Q$.

The evolution of the system is similar to that of the gravitational case in
Sect.\,\ref{sss.Gravity}, however, attraction and repulsion have to be
considered during the approach or separation of bodies C and D. The initial
distance between C and D be $r$, when both bodies are assumed to be at rest,
and changes to $r \pm \Del r$ by
the repulsive or attractive force~$K_{\rm E}(r)$ between
the charges given by Coulomb's law in Eq.\,(\ref{eq.Coulomb})
The force is approximately constant for $r \gg \Del r > 0$
causing accelerations of $b_{\rm D} = K_{\rm E}(r)/m_{\rm D}$ and
$b_{\rm C} = - K_{\rm E}(r)/m_{\rm C}$, respectively.
Since the duration~$\Del t$ of
the motions of both bodies is the same, the separation (upper sign) or
approach (lower sign) of C and D can be formulated as follows:
%
%% Eq. 66
\begin{eqnarray}
\pm \Del r = \pm (s_{\rm D} - s_{\rm C}) =
\pm \frac{1}{2}\,(b_{\rm D} - b_{\rm C})\,(\Del t)^2 = \nonumber \\
\frac{1}{2}\,\left(\frac{1}{m_{\rm C}} +
\frac{1}{m_{\rm D}}\right)\,K_{\rm E}(r)\,(\Del t)^2 ~.%D\_delta\_r
\label{eq.D_delta_r}
\end{eqnarray}
Comparing the second term of the equation with the last one, it can be seen that
$s_{\rm D}\,m_{\rm D} = - s_{\rm C}\,m_{\rm C}$, i.e. the
centre of gravity stays at rest.
Multiplication of Eq.\,(\ref{eq.D_delta_r}) by~$K_{\rm E}(r)$
gives a good estimate of the corresponding kinetic energy:
%
%% Eq. 67
\begin{eqnarray}
\pm \Del r\,K_{\rm E}(r) =
\pm \Del r\,\frac{Q\,q}{4\,\pi\,\varepsilon_0\,r^2} = \nonumber \\
\frac{1}{2}\,\frac{K_{\rm E}^2(r)}{m_{\rm B}}\,(\Del t)^2 +
\frac{1}{2}\,\frac{K_{\rm E}^2(r)}{m_{\rm A}}\,(\Del t)^2 = \nonumber \\
\frac{1}{2}\,m_{\rm D}\,v_{\rm D}^2 +
\frac{1}{2}\,m_{\rm C}\,v_{\rm C}^2 = T_{\rm D} + T_{\rm C} > 0 ~,% acc
\label{eq.acc}
\end{eqnarray}
where $v_{\rm D} = b_{\rm D}\,\Del t$ and $v_{\rm C} = b_{\rm C}\,\Del t$
are the speeds of the bodies, when the
distances~$r \pm \Del r$ between C and D are attained.
The sum of the kinetic energies $T_{\rm C}$ and $T_{\rm D}$
must, of course, be equal to the difference of the electrostatic
potential energy at distances of $r$ and $r \pm \Del r$:
%
%% Eq. 68
\begin{eqnarray}
\{\Ph(r) - \Ph(r \pm \Del r)\}\,q = \nonumber \\
\frac{Q\,q}{4\,\pi\,\varepsilon_0}\,\left(\frac{1}{r} - \frac{1}
{r \pm \Del r} \right) \approx
\pm \Del r\,\frac{Q\,q}{4\,\pi\,\varepsilon_0\,r^2} > 0 ~.% diff\_pot
\label{eq.diff_pot}
\end{eqnarray}
The variations of the number of $\Del N_{Q,q}(r)$ dipoles in
Eqs.\,(\ref{eq.number}) and (\ref{eq.pot_energy}) during the
separation or approach of bodies C and D from $r$ to $r \pm \Del r$ are
%
%% Eq. 69
\begin{eqnarray}
\del N_{Q,q}(r,\Del r) =
\Del N_{Q,q}(r \pm \Del r) - \Del N_{Q,q}(r) = \nonumber\\
\frac{\eta_{\rm D}\,\kappa_{\rm D}}{c_0^2}\,\frac{Q\,|q|}
{4\,\pi}\left[\frac{1}{r \pm \Del r} - \frac{1}{r}\right] \approx
\mp \Del r\,\frac{\eta_{\rm D}\,\kappa_{\rm D}}{c_0^2}\,\frac{Q\,|q|}
{4\,\pi\,r^2} ~.% ~num
\label{eq.num}
\end{eqnarray}
The number of reversed dipoles thus decreases during the
separation of~C and D in Fig.~\ref{fig:plus}.
The corresponding energy variation with positive $q$ is,
cf. Eq.\,(\ref{eq.pot_energy}):
%
%% Eq. 70
\begin{eqnarray}
\del E_{Q,q}(r,\Del r) = 2\,p_{\rm D}\,c_0\,\del N_{Q,q}(r,\Del r) = \nonumber \\
- \Del r\,|q|\,\frac{Q}{4\,\pi\,\varepsilon_0\,r^2} < 0 ~.% D\_Num\_en
\label{eq.D_Num_en}
\end{eqnarray}
The energy of the reversed dipoles thus decreases
by the amount that fuels the kinetic energy in Eq.\,(\ref{eq.acc}).

In the opposite case with negative~$q$ and attraction, it
can be seen from Fig.~\ref{fig:minus} that the increased number of reversed
dipoles is actually leaving the system without momentum exchange and is lost.
The momentum difference, therefore, is again negative
%
%% Eq. 71
\begin{equation}
\del P_{Q,q}(r,\Del r) = - 2\,p_{\rm D}\,\del N_{Q,q}(r,\Del r) %D\_diff\_mon
\label{eq.D_diff_mon}
\end{equation}
and so is the energy of the reversed dipoles confined in the system:
%
%% Eq. 72
\begin{eqnarray}
\del E_{Q,q}(r,\Del r) = \del P_{Q,q}(r,\Del r)\,c_0 = \nonumber \\
- |q|\,\Del r\,\frac{Q}{4\,\pi\,\varepsilon_0\,r^2} < 0 ~.% D\_diff\_en
\label{eq.D_diff_en}
\end{eqnarray}

The electrostatically bound two-body system thus is a closed system
in the sense defined by \citet{Lau20}, slowly evolving in time during the
movements of the bodies~C and D. The potential energy converted into kinetic
energy stems from the modified dipole distributions.

%%%%%%%%%%%%%%%%%%%%%%%%%%%%%%%%%%%%%%%%%%%%%%%%%%%%%%%%%%%%%%%%%%%%%%%%%%%%%%
%%%%%%%%%%%%%%%%%%%%%%%%%%%%%%%%%%%%%%%%%%%%%%%%%%%%%%%%%%%%%%%%%%%%%%%%%%%%%%

\subsection{Pioneer anomaly}
\label{ss.Pioneer}
%% Subsect. 3.2

Anomalous frequency shifts of the Doppler radio-tracking signals were
detected for both Pioneer spacecraft \citep{And98}. The observations of
Pioneer~10 (launched on 2~March 1972) published by the Pioneer Team will be
considered during the time interval, $(t_0, t_1)$, between 3~January 1987 and
22~July 1998 ($t_1 - t_0 \approx 11.55\,{\rm years} = 3.645 \times 10^8\s$),
while the spacecraft was at heliocentric distances between $r_0 = 40$\,ua and
$r_1 = 70.5$\,ua. The Pioneer team took into account all known
contributions in calculating a model frequency, $\nu_{\rm mod}(t)$, which was
based on a constant clock frequency~$f_0$ at the terrestrial control stations.
Observations at times $t = t_0 + \Del t$ then indicated a nearly uniform
increase of the observed frequency shift with respect to the expected one of
%
%% Eq. 73
\begin{equation}
\nu_{\rm obs}(t) - \nu_{\rm mod}(t) = 2\,\dot{f}\,\Del t %  ~eq.anomaly
\label{eq.anomaly}
\end{equation}
with $\dot{f} = 5.99 \times 10^{-9}\Hz\s^{-1}$ \citep[cf.][]{Tur06}.

The observations of the anomalous frequency shifts
could, in principle, be interpreted as a deceleration of the
heliocentric spacecraft velocity by
%
%% Eq. 74
\begin{equation}
a_{\rm p} = - (8.74 \pm 1.33) \times 10^{-10}\m\s^{-2}~.
\label{eq.deceleration}
\end{equation}
However, no unknown sunward-directed force could be identified
\citep[cf.][]{Ior07}. Alternatively, a clock acceleration at the ground
stations of
%
%% Eq. 75
\begin{equation}
a_t = \frac{a_{\rm p}}{c_0} = (2.92 \pm 0.44) \times 10^{-18}\s^{-1}
\label{eq.advanced}
\end{equation}
could explain the anomaly. A true trajectory anomaly together with an unknown
systematic spacecraft effect was considered to be the most likely
interpretation by \citet{And02}. Although \citet{Tur12} later concluded that
thermal recoil forces of the spacecraft caused the anomaly of Pioneer~10, the
discussion in the literature continued.

Assuming an atomic clock acceleration, a constant reference frequency~$f_0$
for the calculation of $\nu_{\rm mod}(t)$ is not appropriate
\citep[cf.][]{WilDwi11}. Consequently the
equation
%
%% Eq. 76
\begin{equation}
[\nu_{\rm obs}(t) - f_0] -[\nu_{\rm mod}(t) - f_0]  = 2\,\dot{f}\,\Del t ~,
\label{eq.anomaly2}
\end{equation}
equivalent to Eq.\,(\ref{eq.anomaly}), has to be modified with
%
%% Eq. 77
\begin{equation}
f(t) = f_0 + \dot{f}\,\Del t = f_0\,(1 + \frac{\dot{f}}{f_0}\,\Del t) =
f_0\,(1 + a_t\,\Del t) % ~acc\_clock
\label{eq.acc_clock}
\end{equation}
and
%
%% Eq. 78
\begin{equation}
\nu^*_{\rm mod}(t) = \nu_{\rm mod}(t) + 2\,\dot{f}\,\Del t % ~mod\_mod
\label{eq.mod_mod}
\end{equation}
to
%
%% Eq. 79
\begin{eqnarray}
[\nu_{\rm obs}(t) - (f_0 + \dot{f}\,\Del t)] -
[\nu^*_{\rm mod}(t) - (f_0 + \dot{f}\,\Del t)] \nonumber \\ =
[\nu_{\rm obs}(t) - f(t)] -
[\nu^*_{\rm mod}(t) - f(t)] = 0 ~. % no\_anomaly
\label{eq.no_anomaly}
\end{eqnarray}

The gravitationally impact model in Sect.\,\ref{ss.Newton} leads to a secular
mass increase of massive particles in Eq.\,(\ref{eq.G_growth}).
Consequently the Rydberg constant in Eq.\,(\ref{eq.Rydberg}) would increase
in a linear approximation with the electron mass~$m_{\rm e}$ according to
%
%% Eq. 80
\begin{equation}
R^*_\infty(t) = \frac{\alpha^2\,c_0}{2\,h}\,m_{\rm e}\,(1 + A\,\Del t) %~
%Rydberg\_increased
\label{eq.Rydberg_increased}
\end{equation}
resulting in frequency increases of atomic clocks with time. They give rise to
the clock acceleration in Eq.\,(\ref{eq.acc_clock}), if we assume $a_t = A$.
The most likely values of $r_{\rm G,e}$ in Fig.~\ref{fig:G_en_Y_H} range from
$2.04 \times 10^{-4}\pmeter$ to 2.82~fm, the classical electron radius,
corresponding with Eq.\,(\ref{eq.G_accretion}) to
$A_{\rm H} = 2.43 \times 10^{-18}\s^{-1} = H_0$, the Hubble constant, and
$A \approx 1.3 \times 10^{-20}\s^{-1}$.
Within the uncertainty margins the high value agrees with $a_t$ in
Eq.\,(\ref{eq.advanced}) and would
quantitatively account for the Pioneer frequency shift.

Should the anomaly be much less pronounced, because thermal recoil forces
decelerate the spacecraft, the range of $A$ in Fig.~\ref{fig:G_en_Y_H} could
thus accommodate smaller values of $a_t$ as well.

%%%%%%%%%%%%%%%%%%%%%%%%%%%%%%%%%%%%%%%%%%%%%%%%%%%%%%%%%%%%%%%%%%%%%%%%%%%%%%
%%%%%%%%%%%%%%%%%%%%%%%%%%%%%%%%%%%%%%%%%%%%%%%%%%%%%%%%%%%%%%%%%%%%%%%%%%%%%%

\subsection{Sun-Earth distance increase}
\label{ss.Earth_Sun}
%% Subsect. 3.3

A secular increase of the mean Sun-Earth distance with a rate
of $(15 \pm 4)\m$ per century had been reported using
many planetary observations between 1971 and 2003
%(corresponding to $1 \times 10^9\s$)
\citep{KraBru}.
Neither the influence of cosmic expansion nor a time-dependent
gravitational constant seem to provide an explanation \citep{Lae08}.

As our impact model summarized in Sect.\,\ref{ss.Newton} leads to a secular
mass increase according to Eq.\,(\ref{eq.G_growth})
of all massive bodies fuelled by
a decrease in energy of background flux of gravitons, it allowed us to
formulate a quantitative understanding of the effect within the parameter
range of the model \citep{WilDwi13b}.

The value of the astronomical unit is now defined as
$1\ua = 1.495\,978\,707\,00 \times 10^{11}\m$ (exact) by the
International Astronomical Union (IAU) and the
Bureau International des Poids et Mesure \citep{BIPM}.
The mean Sun-Earth distance is known with a standard uncertainty
of (3 to 6)\,m for $r_{\rm E}$ \citep{PitSta,AndNie,HarPrs}.

Considering this uncertainty, the measurement of a change rate of
%
%% Eq. 81
\begin{equation}
\frac{\Del r_{\rm E}}{\Del t} = \frac{(15 \pm 4)\m}{3.156 \times 10^9\s} =
(4.8 \pm 1.3)\nm\s^{-1} % ~eq.change
\label{eq.change}
\end{equation}
is difficult, but feasible as relative determination.
A circular orbit approximation had been considered, because the mean value
of $r_{\rm E}$ is of interest:
%
%% Eq. 82
\begin{equation}
r_{\rm E} = \frac{G_{\rm N}\,M_\odot}{v^2_{\rm E}} =
\frac{\mu_\odot}{v^2_{\rm E}}~. % eq.Kepler\_mod
\label{eq.Kepler_mod}
\end{equation}
following from Eq.\,(\ref{eq.Newton}) and the centrifugal force with
$v_{\rm E}$, the tangential orbital velocity of the Earth, where
the heliocentric gravitational constant is\\
$\mu_\odot = 1.327\,124\,400\,41 \times 10^{20}\m^3\s^{-2}$ (IAU) and thus
the mass of the Sun
$M_\odot = (1.988\,42 \pm 0.000\,25) \times 10^{30}\kg$.

We now consider Eq.~(\ref{eq.Kepler_mod}) not only for $t_0$, but also at
$t = t_0 + \Del t$ assuming constant $G_{\rm N}$ as well as constant
$v_{\rm E}$. The
latter assumption is justified by the fact that any uniformly moving particle
does not experience a deceleration. It implies an increase of the momentum
together with the mass accumulation of the Earth. The apparent violation of
the momentum conversation principle can be resolved by considering the
accompanying momentum changes of the graviton distribution. A detailed
discussion of this aspect is given in Sect.~3 of Paper\,1.

From Eqs.~(\ref{eq.G_growth}) and (\ref{eq.Kepler_mod}) it
followed
%
%% Eq. 83
\begin{equation}
r_{\rm E}(t) = r_{\rm E} + \Del r_{\rm E} \approx
\frac{G_{\rm N}}{v^2_{\rm E}}\,M_\odot\,(1 + A\,\Del t) % ~eq.equivalent
\label{eq.equivalent}
\end{equation}
and
%
%% Eq. 84
\begin{equation}
\frac{\Del r_{\rm E}}{\Del t} \approx r_{\rm E}\,A ~. % eq.Del\_r
\label{eq.Del_r}
\end{equation}
With the help of Eqs.~(\ref{eq.G_accretion}) and (\ref{eq.change}),
the electron mass radius can now be calculated. The result is
%
%% Eq. 85
\begin{eqnarray}
r_{\rm G,e} =
\left(r_{\rm E}\,\frac{\Del t}
{\Del r_{\rm E}}~1.014 \times 10^{-49}\m^2\s^{-1}\right)^{-1/2} =
\nonumber \\
\left(1.8^{+0.4}_{-0.2}\right)\fmeter ~,% ~eq.r\_G\_e
\label{eq.r_G_e}
\end{eqnarray}
close to the classical electron radius
%
%% Eq. 86
\begin{equation}
r_{\rm e} = \alpha^2\,a_0 = 2.82\fmeter ~.
\label{electron_radius}
\end{equation}
The relative accumulation rate deduced from
the observations of $r_{\rm E}$ finally becomes \\ $A = A_{\rm ua} \approx
3.2\times 10^{-20}\s^{-1}$ (see Fig.\,\ref{fig:G_en_Y_H}).

%%%%%%%%%%%%%%%%%%%%%%%%%%%%%%%%%%%%%%%%%%%%%%%%%%%%%%%%%%%%%%%%%%%%%%%%%%%%%%
%%%%%%%%%%%%%%%%%%%%%%%%%%%%%%%%%%%%%%%%%%%%%%%%%%%%%%%%%%%%%%%%%%%%%%%%%%%%%%

\subsection{Secular perihelion advances in the solar system}
\label{ss.perihel}
%% Subsect. 3.4

Multiple application of the interaction process described in
Sect.\,\ref{ss.Newton} can
produce gravitons with reduction parameters greater than $Y$ in large mass
conglomerations\,--\,within the Sun in this section. The proportionality of
the linear term in the binomial theorem with the exponent~$n$ in
%
%% Eq. 87
\begin{equation}
(1 - Y)^n \approx 1 - n\,Y~~~~~{\rm for}~~Y \ll 1 % ~eq.binom
\label{eq.binom}
\end{equation}
suggests that a linear superposition of the effects of multiple interactions
will be a good approximation, if~$n$ is not too large. Energy reductions
according to Eq.\,(\ref{eq.G_reduction}) are therefore not lost, as claimed
by \citet{Dru97}, but they are redistributed to other emission locations within
the Sun. This has two consequences: (1) The total energy reduction is still
dependent on the solar mass, and (2) since emissions from matter closer to the
surface of the Sun in the direction of an orbiting object is more likely to
escape into space than gravitons from other locations, the effective
gravitational centre should be displaced from the centre of the Sun towards
that object.

Using published data on the secular perihelion advances of the inner planets
Mercury, Venus, Earth and Mars of the solar system and the asteroid Icarus,
we found that
the effective gravitational centre is displaced from the centre of the Sun by
approximately $\rho = 4400\m$ \citep{WilDwi14b}. Since an analytical derivation of this value
from the mass distribution of the Sun was beyond the scope of the study,
future investigations need to show that the
modified process with directed secondary graviton emission can quantitatively
account for such a displacement.

%%%%%%%%%%%%%%%%%%%%%%%%%%%%%%%%%%%%%%%%%%%%%%%%%%%%%%%%%%%%%%%%%%%%%%%%%%%%%%
%%%%%%%%%%%%%%%%%%%%%%%%%%%%%%%%%%%%%%%%%%%%%%%%%%%%%%%%%%%%%%%%%%%%%%%%%%%%%%

\subsection{Planetary flyby anomalies}
\label{ss.flyby}
%% Subsect. 3.5
%%%%%%%%%%%%%%%%%%%%%%%%%%%%%%%%%%%%%%%%%%%%%%%%%%%%%%%%%%%%%%%%%%%%%%%%%%%%%%
\subsubsection{Earth flybys}
\label{sss.Earth_flyby}
%% Subsubsect. 3.5.1

Several Earth flyby manoeuvres indicated anomalous accelerations and
decelerations and led to many investigations without reaching a solution of
the problem, see recent reviews by \citet{And08} and \citet{NieAnd}.
Since there is general agreement that the anomaly is only significant
near perigee, we discuss here the seven passages at altitudes below
2000\,km listed in Table\,1 of \citet[][note the wrong dates]{Ace17}. %3
Three of them (Galileo\,I, NEAR and Rosetta) we have studied
assuming the gravitational impact model of Sect.\,\ref{ss.Newton} and
multiple interactions \citep{WilDwi15c}. As in Sect.\,\ref{ss.perihel},
the multiple interactions result in a deviation~$\rho$ of the effective
gravitational centre from the geometric centre. We obtained for Galileo,
NEAR and Rosetta $\rho \approx 1.3\m,~3.9\m~{\rm and}~0.5\m$,
respectively. The study had been conducted assuming a spherically symmetric
emission of liberated gravitons mentioned in Sect.\,\ref{ss.Newton}.

With the assumption of an anti-parallel emission, we have repeated the
analysis and found $\rho \approx 2\m$ for all spacecraft, provided
the origin of $\rho$ is shifted by approximately $-0.6\m$ in the direction
of the perigee of Galileo\,I, $+1.9\m$ for NEAR, and $-1.5\m$ for Rosetta.
Moreover, it was possible to model the decelerations of Galileo\,II on
8 December 1992 with a shift of $-3.4\m$; of Cassini on 18 August 1999 with
$-2.7\m$ and the null result for Juno on 9 October 2013 with $-2\m$.

An origin offset of $+3.4\m$ opposite to the Cassini perigee could to a first
approximation achieve all apparent shifts taking the geographic coordinates
of the various flybys into account. A detailed study would have to consider
in addition the Earth gravitational model.

%%%%%%%%%%%%%%%%%%%%%%%%%%%%%%%%%%%%%%%%%%%%%%%%%%%%%%%%%%%%%%%%%%%%%%%%%%%%%%
\subsubsection{Juno Jupiter flybys}
\label{sss.Jupiter_flyby}
%% Subsubsect. 3.5.2

Juno was inserted into an elliptical orbit around Jupiter on 4 July 2016 with
an orbital period of 53.5~days. \citet{Aceetal} studied the first and the
third orbit with a periapsis of ``4200\,km over the planet top clouds''.
``A significant radial component was found and this decays with the distance
to the center of Jupiter as expected from an unknown physical interaction.
...
The anomaly shows an asymmetry among the incoming and outgoing branches
of the trajectory ... .''
The radial component is shown in their Figure~6 between
$t = (-180~{\rm to}~+180)$\,min around perijove for the first and third
Juno flyby. The peak anomalous outward accelerations shown are in both cases:
$\delta a = 7\mm\s^{-2}$ at $t \approx -15$\,min and $\delta a = 6\mm\s^{-2}$
at $t \approx +17$\,min.

\begin{figure}[!t]
% Figure 7
\centering
\includegraphics[width=\columnwidth]{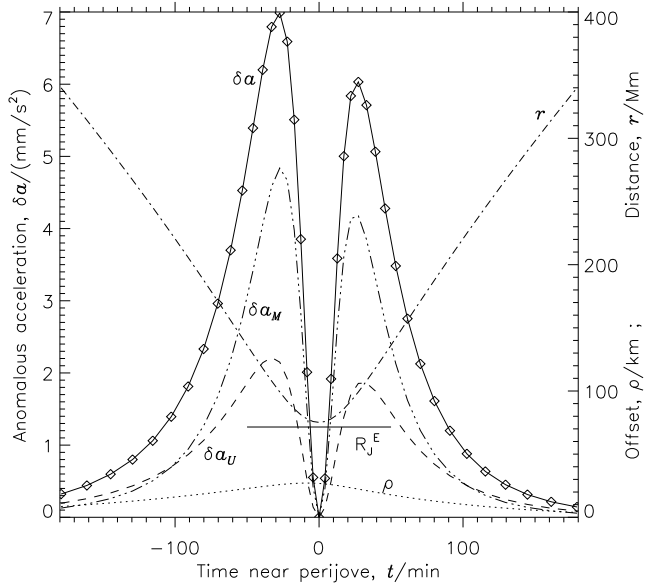}
\caption{Anomalous radial outward acceleration~$\delta a$ experienced by Juno
near perijove at time~$t = 0$ (solid curve with diamond signs).
It is composed of $\delta a_U$ calculated from the adjusted potential
and $\delta a_M$ calculated from the adjusted centrifugal energy
\citep[cf.][see effective potential energy equation~14.8]{LanLif}.
A multi-interaction process has been assumed within the  mass
$1.898\,58 \times 10^{27}\kg$ of Jupiter. It causes an offset~$\rho$ of the
effective pivotal point of the gravitational attraction from the geometric
centre of Jupiter (dotted curve). Also shown are the equatorial radius of
Jupiter~${R_J}^E$ (solid bar) and the radial distance~$r$ of Juno from the
centre (dash-dot curve).}
\label{fig:Juno}
\end{figure}

We applied the multiple-interaction concept of the previous
Sects.\,\ref{ss.perihel} and \ref{sss.Earth_flyby}, and found that offsets
of $\rho \approx$ (8 to 27)\,km of the gravitational from the geometric centre
are required to model the acceleration in Fig.~\ref{fig:Juno}, which is in
good agreement with the observations during the Juno Jupiter flybys.
The variation of $\rho$ could be modelled by an ellipsoidal displacement
of the gravitational centre offset in the direction of a flyby position near
$t = -10$\,min.
%

%%%%%%%%%%%%%%%%%%%%%%%%%%%%%%%%%%%%%%%%%%%%%%%%%%%%%%%%%%%%%%%%%%%%%%%%%%%%%%
%%%%%%%%%%%%%%%%%%%%%%%%%%%%%%%%%%%%%%%%%%%%%%%%%%%%%%%%%%%%%%%%%%%%%%%%%%%%%%

\subsection{Rotation velocities of spiral galaxies}
\label{ss.galaxies}
%% Subsect. 3.5

The rotation velocities of spiral galaxies are difficult to
reconcile with Keplerian motions, if only the gravitational effects of the
visible matter is taken into account \citep[e.g.][]{Rub83,Rub86}. Dark matter
had been proposed by \citet{Oor32} and \citet{Zwi33} in order to understand
several velocity anomalies in galaxies and clusters of galaxies.
A MOdification of the Newtonian Dynamics (MOND) has been introduced by
\citet{Mil83} that assumes a modified gravitational interaction at low
acceleration levels.

The impact model of gravitation
in Sect.\,\ref{ss.Newton} is applied to the radial acceleration of disk
galaxies \citep{WilDwi18}. The flat velocity curves of NGC\,7814, NGC\,6503
and M\,33 are obtained without the need to postulate any dark matter
contribution. The concept explained below provides a physical process that
relates the fit parameter of the acceleration scale defined by \citet{McGetal}
to the mean free path length of gravitons in the disks of galaxies. It may
also provide an explanation for MOND.

\citet{McG05} has observed a fine balance between baryonic and dark mass in
spiral galaxies that may point to new physics for DM or a modification of
gravity. \citet{FrSaKa} have also concluded that either the baryons dominate
the DM or the DM is closely coupled with the luminous component.
\citet{SalTur} have suggested that there is a profound interconnection
between the dark and the stellar components in galaxies.

The large baryonic masses in galaxies will cause multiple interactions of
gravitons with matter if their propagation direction is within the disk. For
each interaction the energy loss of the gravitons is assumed to be
$Y\,T_{\rm G}$ (for details see Sect.\,2.3 of Paper\,1). The important
point is that the multiple interactions occur only in the galactic plane and
not for inclined directions. An interaction model is designed
indicating that an amplification factor of approximately two can be achieved
by six successive interactions.  An amplification
occurs for four or more interactions. The process works, of course, along each
diameter of the disk and leads to a two-dimensional distribution of reduced
gravitons.

The multiple interactions do not increase the
total reduction of graviton energy, because the number of interactions
is determined by the (baryonic) mass of the gravitational centre
according to Paper\,1. A galaxy with enhanced gravitational acceleration
in two dimensions defined by the galactic plane, will, therefore, have a
reduced acceleration in directions inclined to this plane.

%%%%%%%%%%%%%%%%%%%%%%%%%%%%%%%%%%%%%%%%%%%%%%%%%%%%%%%%%%%%%%%%%%%%%%%%%%%%%%
%%%%%%%%%%%%%%%%%%%%%%%%%%%%%%%%%%%%%%%%%%%%%%%%%%%%%%%%%%%%%%%%%%%%%%%%%%%%%%

\subsection{Light deflection and Shapiro delay}
\label{ss.Shapiro}
%% Subsect. 3.7

The deflection of light near gravitational centres is of fundamental
importance. For a beam passing close to the Sun \citet{Sol04} and \citet{Ein11}
obtained a deflection angle of $0.87\arcsec$ under the assumption
that radiation would be affected in the same way as matter.
\emph{Twice} this value was then derived in the framework of the GTR
\citep{Ein16}\footnote{It is of interest in
the context of this paper that Einstein employed Huygens' Principle in his
calculation of the deflection.},
and later by \citet{Sch60} using the equivalence principle and STR.
The high value was confirmed
during the total solar eclipse in 1919 for the first time \citep[]{Dysetal}.
This and later observations have been summarized by \citet{Mik59} and
combined to a mean value of
approximately 2\arcsec.

The deflection of has also been considered in the context of the
gravitational impact model summarized in Sect.\,\ref{ss.Newton}. As
a secular mass increase of matter was a consequence
of the this model, the question arises of how the interaction of gravitons
with photons can be understood, since the photon mass is in all likelihood
zero.\footnote{A zero mass of photons follows from the STR and a speed of light
in vacuum~$c_0$ constant for all frequencies. \citet{Ein05a} used
,,Lichtquant'' for a quantum of electromagnetic radiation;
the term ``photon'' was introduced by \citet{Lew26}.
With various methods the photon mass could be constrained to
$m_\nu < 10^{-49}\kg$ \citep{GolNie,Amsetal} or even to
$m_\nu < 6.3 \times 10^{-53}\kg$ \citep{YanZha}.}
An initial attempt at solving that problem has been made in
\citet{WilDwi13a} by assuming that a
photon stimulates an interaction with a rate equal to its frequency
$\nu = E_\nu/h$. It is summarized here under the assumption of an
anti-parallel re-emission, both for massive particles and photons.

A physical process will then be outlined that provides information on
the gravitational potential~$U$ at the site of a photon emission
\citep{WilDwi17}. This aspect
had not been covered in our earlier paper on the gravitational redshift
\citep{WilDwi14a}.

Interactions between massive bodies have been treated in Paper~1 with an
absorption rate of \emph{half} the intrinsic de Broglie frequency of a mass,
because \emph{two} virtual gravitons have to be emitted for one interaction.
The momentum transfer to a photon will thus be twice
as high as to a massive body with a mass equivalent to $E_\nu/c^2_0$.

We then apply the momentum conservation principle to photon-graviton pairs in
the same way as to photons \citep[cf.][]{LanLif} and can write after a
reflection of $\vec {p}_{\rm G}$
%
%% Eq. (88)
\begin{equation}
\vec {p}_\nu + \vec {p}_{\rm G} =
\vec {p}_\nu + 2\,\vec {p}_{\rm G} - \vec {p}_{\rm G} =
\vec {p}^*_\nu - \vec {p}_{\rm G} % ~ eq.photon\_momentum
\label{eq.photon_momentum}
\end{equation}
with $|{\vec p}_{\rm G}| = p_{\rm G} = T_{\rm G}/c_0$.

We assume, applying Eq.\,(\ref{eq.photon_momentum}) with
$p_{\rm G} \ll p_\nu = |\vec{p_\nu}|$,
that under the influence of a gravitational centre relevant
interactions occur on opposite sides of a photon with $p_{\rm G}$ and
$p_{\rm G}\,(1 - Y)$ transferring a net momentum of $2\,Y\,p_{\rm G}$. Note,
in this context, that the Doppler effect can only operate for interactions of
photons with massive bodies \citep[cf.][]{Fer32,Som78}.
Consequently, there will be no energy change of the photon, because both
gravitons are reflected with constant energies under these
conditions, and we can write for a pair of interactions
%
%% Eq. (89)
\begin{equation}
E_\nu = |\vec p_\nu|\,c = |\vec p_\nu + 2\,Y\vec p_{\rm G}|\,c' =
|\vec p'_\nu|\,c' = E'_\nu ~, % eq.photon\_energy
\label{eq.photon_energy}
\end{equation}
where $\vec p'_\nu$ is the photon momentum after the events. If $\vec p_\nu$
and a component of $2\,Y\vec p_{\rm G}$ are pointing in the same direction,
it is $c' < c$, the speed is reduced; an antiparallel direction leads to
$c' > c$. Note that this could, however, not result in $c' > c_0$, because
$c = c_0$ can only be attained in a region with
an isotropic distribution of gravitons with a momentum of $p_{\rm G}$,
i.e. with a gravitational potential~$U_0 = 0$.

The momentum $\vec p_\nu$ of a photon radially approaching a gravitational
centre will be treated in line with Eq.\,(6) of Paper~2 for massive bodies,
however, with twice the interactionrate for photons.
Since we know from observations that the deflection
of light near the Sun is very small, the
momentum variation caused by the weak and static
gravitational interaction is also very small.
The momentum change rate of the photon can then be approximated by
%
%% Eq. (90)
\begin{equation}
\frac{\Del {\vec p}_\nu}{\Del t} \approx
2\,G_{\rm N}\,M_\odot\,\frac{\uvec r}{r^2}\,\frac{p_\nu}{c_0} ~,%eq.transfer\_rate
\label{eq.transfer_rate}
\end{equation}
where  $r = |\vec r|$ is the
distance of the photon from the centre, and the position vector of
the photon is $r\,\uvec{r}$ with a  unit vector $\uvec{r}$.
The small deflection angle also allows an approximation of
the actual path by a straight line along an $x$~axis: $x \approx c_0\,t$.
The normalized momentum variation along the trajectory then is
%
%% Eq. (91)
\begin{equation}
-\frac{c_0}{p_\nu}\left(\frac{\Del \vec{p}_\nu}{\Del t}\right)_x =
\frac{c_0}{p_\nu}\,\frac{\Del p_\nu}{\Del t}\cos \vartheta
\approx 2\,G_{\rm N}\,M_\odot\,\frac{x}{r^3}~. % eq.x\_component
\label{eq.x_component}
\end{equation}
The corresponding component perpendicular to the trajectory is
%
%% Eq. (92)
\begin{equation}
-\frac{c_0}{p_\nu}\left(\frac{\Del \vec{p}_\nu}{\Del t}\right)_y =
\frac{c_0}{p_\nu}\,\frac{\Del p_\nu}{\Del t}\,\sin \vartheta
\approx 2\,G_{\rm N}\,M_\odot\frac{R}{r^3}~,%eq.y\_component
\label{eq.y_component}
\end{equation}
where $R$ is the impact parameter of the trajectory.
Integration of Eq.\,(\ref{eq.x_component}) over $t$ from $-\infty$ to $x/c_0$
yields
%
%% Eq. (93)
\begin{equation}
\frac{1}{p_\nu}\,[\Del \vec{p}_\nu(r)]_x \approx
\frac{2\,G_{\rm N}\,M_\odot}{c_0^2\,r} =
\frac{2\,G_{\rm N}\,M_\odot}{c_0^2\,\sqrt{R^2 + x^2}}~.%eq.x\_integrated
\label{eq.x_integrated}
\end{equation}

If we apply Eq.\,(\ref{eq.photon_energy}) to a photon approaching the
Sun along the $x$~axis
starting from infinity with $E_\nu = p_\nu\,c_0$, and
considering that the $y$~component in Eq.\,(\ref{eq.x_component}) is much smaller
than the x component in Eq.\,(\ref{eq.y_component}) for $x \gg R$,
the photon speed~$c(r)$ as a function of $r$
can be determined from
%
%% Eq. (94)
\begin{equation}
p_\nu\,c_0 \approx \{p_\nu + [\Del \vec{p}_\nu(r)]_x\}\,c(r)~.% eq.reduced\_c
\label{eq.reduced_c}
\end{equation}
Division by $p_\nu\,c_0$ then gives with Eq.\,(\ref{eq.x_integrated})
%
%% Eq. (95)
\begin{equation}
\frac{1}{[n_{\rm G}(r)]_x} = \frac{c(r)}{c_0} \approx
1 - \frac{2\,G_{\rm N}\,M_\odot}{c_0^2~r} = 1 + \frac{2\,U(r)}{c_0^2}%~eq.refraction
\label{eq.refraction}
\end{equation}
as a good approximation of the inverse gravitational index of refraction
along the $x$~axis. The same index has been obtained albeit with different
arguments, e.g. by \citet{Booetal,YeLin}. The resulting speed of light is in
agreement with evaluations by \citet{Sch60}, for a radial
propagation\footnote{\citet{Ein12} states
explicitly that the speed at a certain location is not dependent on the
direction of the propagation.} in a central gravitational field, and
\citet{Oku00}\,---\,calculated on the basis of the standard Schwarzschild
metric. A decrease of the speed of light near the Sun, consistent with
Eq.\,(\ref{eq.refraction}), is not only supported by the predicted and
subsequently observed Shapiro delay
\citep{Sha64,Reaetal,Sha71,Kraetal,Baletal,KutZaj},
but also indirectly by the deflection of light \citep{Dysetal}.

The deflection of light by gravitational centres according to the
GTR \citep{Ein16} and its observational detection
by \citet{Dysetal} leave no doubt that a photon is deflected by a factor of two
more than expected relative to a corresponding massive particle. Since in
our concept the interaction rate between photons and gravitons is twice as
high as for massive particles of the same total energy, the reflection of
a graviton from a photon with a momentum of $(1 - Y)\,p_{\rm G}$ must also
be anti-parallel to the incoming one, i.e. a momentum
of~$- 2\,Y\,p_{\rm G}$ will be transferred. Otherwise the correct deflection
angle for photons cannot be obtained.
This modified interaction process has one further important advantage: the
reflected graviton can interact with the deflecting gravitational centre
and\,--\,through the process outlined in the paragraph just before
Eq.\,(\ref{eq.imbalance})\,--\,transfers~$2\,Y\,p_{\rm G}$, in compliance with
the momentum conservation principle. In the old scheme, the violation of this
principle had no observational consequences, because of the extremely large
masses of relevant gravitational centres, but the adherence to both the
momentum and energy conservation principles is very encouraging and clearly
favours the new concept.

Basically the same arguments are relevant for the longitudinal interaction
between photons and gravitons. The momentum transfer per interaction will
be doubled, but the gravitational absorption coefficient will be reduced
by a factor of two. Together with an increased graviton density, all
quantities and results are the same as before. However, a detailed analysis
shows that the momentum conservation principle is now also adhered to.

%%%%%%%%%%%%%%%%%%%%%%%%%%%%%%%%%%%%%%%%%%%%%%%%%%%%%%%%%%%%%%%%%%%%%%%%%%%%%%
%%%%%%%%%%%%%%%%%%%%%%%%%%%%%%%%%%%%%%%%%%%%%%%%%%%%%%%%%%%%%%%%%%%%%%%%%%%%%%

%%%%%%%%%%%%%%%%%%%%%%%%%%%%%%%%%%%%%%%%%%%%%%%%%%%%%%%%%%%%%%%%%%%%%%%%%%%%%%
%%%%%%%%%%%%%%%%%%%%%%%%%%%%%%%%%%%%%%%%%%%%%%%%%%%%%%%%%%%%%%%%%%%%%%%%%%%%%%

\subsection{Gravitational redshift}
\label{ss.redshift}
%% Subsect. 3.8

The gravitational potential~$U$ at a distance~$r$ from a spherical body with
mass~$M$ is constraint in the weak-field approximation for non-relativistic
cases \citep[cf.][]{LanLif} by
%
%% Eq. (96)
\begin{equation}
- 1  \ll \frac{U}{c^2_0} = - \frac{G_{\rm N}\,M}{c^2_0~r} \le 0 ~.%eq.potential
\label{eq.potential}
\end{equation}
A definition of a reference potential in line with
this formulation is $U_\infty = 0$ for $r = \infty$.

The study of the gravitational redshift, predicted for solar radiation by
\citet{Ein08}, is still an important subject in modern physics and
astrophysics \citep[e.g.][]{Kol04,Neg05,Lae09,Pasetal,Tur13}.
This can be exemplified by two conflicting statements. \citet{Woletal} write:
``The clock frequency is sensitive to the
gravitational potential~$U$ and not to the local gravity field
$\vec{g} = \nabla U$.''
Whereas it is claimed by \citet{Mueetal}:
``We first note that no experiment is sensitive to the
absolute potential~$U$.''

Support for the first alternative can be found in many publications
\citep[e.g.][]{Ein08,Lau20,Sch60,Wil74,Okuetal,SinSam}, but it is, indeed,
not obvious how an atom can locally sense the gravitational potential~$U$.
Experiments on Earth, in space and in the Sun-Earth
system~\citep[cf., e.g.][]{PouReb,KraLue,PouSni,Vesetal,LoP91,TakUen} have,
however, quantitatively confirmed in the static weak field approximation a
relative frequency
shift of
%
%% Eq. (97)
\begin{equation}
\frac{\nu - \nu_0}{\nu_0} = \frac{\Del \nu}{\nu_0}
\approx \frac{\Del U}{c^2_0} = \frac{U - U_0}{c^2_0}~,%eq.shift
\label{eq.shift}
\end{equation}
where $\nu_0$ is the frequency of the radiation emitted by a
certain transition at~$U_0$ and $\nu$ the observed frequency there, if the
emission caused by the same transition had occurred at a potential~$U$.

Since Einstein discussed the gravitational redshift and published conflicting
statements regarding this effect, the confusion could still not be cleared up
consistently \citep[cf., e.g.][]{Man06,Sotetal}. In most of his publications
Einstein defined clocks as atomic clocks. Initially he assumed that the
oscillation of an atom corresponding to a spectral line might be an
intra-atomic process, the frequency of which would be determined by the atom
alone \citep{Ein08,Ein11}. \citet{Sco15} also felt that the equivalence
principle and the notion of an ideal clock running independently of
acceleration suggest that such clocks are unaffected by gravity. \citet{Ein16}
later concluded that clocks would slow down near gravitational centres thus
causing a redshift.

The question whether the gravitational redshift is caused by the emission
process (Case~a) or during the transmission phase (Case~b) is nevertheless
still a matter of recent debates. Proponents of (a) are, e.g.:
\citet{Moe57,Craetal,Sch60,Oha76,Okuetal} and of
(b): \citet{Hayetal,Str04,Ran06,Wil06}.

It
is surprising that the same team of experimenters, albeit with different first
authors (Cranshaw et al. and Hay et al.) published different views on the
process of the Pound--Rebka--Experiment. \citet{PouSni} and \citet{Pou00}
pointed out that this experiment could not distinguish between the two
options, because
the invariance of the velocity of the radiation had not been demonstrated.

\citet{Ein17} emphasized that for an elementary emission process not only
the energy exchange, but also the momentum transfer is of importance
\citep[see also][]{Poi00,Abr02,Fer32}. Taking these considerations into
account, we formulated a photon emission process at
a gravitational potential~$U$ \citep{WilDwi14a} assuming that:
\begin{enumerate}
\item [(1)] The atom cannot sense the potential~$U$, in line with the
original proposal by \citet{Ein08,Ein11}, and initially emits the same
energy~$\Del E_0$ at $U > 0$ and $U_0 = 0$.
\item [(2)] It also cannot directly sense the speed of light at the location with a
potential~$U$. The initial momentum thus is $p_0 = \Del E_0/c_0$.
\item [(3)] As the local speed of light is, however, $c(U) \ne c_0$, a photon
having an energy of $\Del E_0$ and a momentum $p_0$ is not able to propagate.
The necessary adjustments of the photon energy and momentum as well as the
corresponding atomic quantities then lead in the interaction region to
a redshift consistent with $h \nu = \Del E_0\,(1 + U/c^2_0)$ and observations.
\end{enumerate}

As outlined in Sect.~\ref{ss.Shapiro}, there is general
agreement in the literature that the local speed of light is
%
%% Eq. (98)
\begin{equation}
c(U) \approx c_0 \left(1 + \frac{2\,U}{c^2_0}\right)%~eq.local\_speed
\label{eq.local_speed}
\end{equation}
in line with Eq.\,(\ref{eq.refraction}) in Sect.\,\ref{ss.Shapiro}. It has,
however, to be noted that the speed~$c(U)$ was obtained for a photon
propagating from $U_0$ to $U$, and, therefore, the physical process which
controls the speed of newly emitted photons at a gravitational potential~$U$
is not yet established.

An attempt to do that will be made by assuming an aether model.
Before we suggest a specific aether model, a few statements on the aether
concept in general should be mentioned. Following \citet{MicMor} famous
experiment, \citet{Ein05b,Ein08} concluded that the concept of a light aether
as carrier of the electric and magnetic forces is not consistent with the STR.
In response to critical remarks by \citet{Wie11}, cf. \citet{Sch90} for
Wiechert's support of the aether, \citet{Lau12} wrote that the existence
of an aether is not a physical, but a philosophical problem, but later
differentiated between the physical world and its mathematical formulation.
A four-dimensional `world' is only a valuable mathematical trick; deeper
insight, which some people want to see behind it, is not involved \citep{Lau59}.

In contrast to his earlier statements, Einstein said at the end of a speech
in Leiden that according to the GTR a space without aether cannot be
conceived \citep{Ein20}; and even more detailed: Thus one could instead of
talking about `aether' as well discuss the `physical properties of space'.
In theoretical physics we cannot do without aether, i.e., a continuum endowed
with physical properties \citep{Ein24}.
\citet{Mic28} confessed at a meeting in Pa\-sa\-de\-na in the
presence of H.A. Lorentz that he clings a little to the aether; and
\citet{Dir51} wrote in a letter to Nature that there are good reasons for
postulating an {\ae}ther.

In Paper\,2 we proposed an impact model for the electrostatic force
based on massless dipoles. The vacuum is thought to be permeated by these
dipoles that are, in the absence of
electromagnetic or gravitational disturbances, oriented and directed randomly
propagating along their dipole axis with a speed of~$c_0$. There is little or
no interaction among them.
We suggest to identify the dipole distribution postulated in
Sect.\,\ref{ss.Coulomb} with an aether.
Einstein's aether mentioned above may, however, be more related to the
gravitational interactions \citep[cf.][]{Gra01}. In this case, we have to
consider the graviton distribution as another component of the aether.

If we assume that an individual dipole interacts with gravitons in the same
way as photons, see Eq.\,(\ref{eq.photon_energy}), according to
%
%% Eq. (99)
\begin{equation}
T_{\rm D} = |\vec p_{\rm D}|\,c = |\vec p_{\rm D} + 2\,Y\vec p_{\rm G}|\,c' =
|\vec p'_{\rm D}|\,c' = T'_{\rm D} ~, % eq.dipole\_energy
\label{eq.dipole_energy}
\end{equation}
where $T_{\rm D}$ and $\vec p_{\rm D}$ refer to the energy and momentum of a
dipole. The condition $p_{\rm D} \gg p_{\rm G}$,
cf. Eq.\,(\ref{eq.photon_momentum}), is fulfilled in the range from
$Y \approx -22~{\rm to}~-15$ for all $r_{\rm D} \le 2.82\fmeter$.

We can then modify
Eqs.\,(\ref{eq.transfer_rate}) to (\ref{eq.reduced_c})
by changing $\nu$ to D and find that Eqs.\,(\ref{eq.refraction}) and
(\ref{eq.local_speed}) are also valid for dipoles with a speed of $c_0$ for
$U_0 = 0$.

Considering that many suggestions have been made to describe photons as
solitons \citep[e.g.][]{Dir27,Vig91,KamSla,Meu13,Ber13,Beretal}, we
also propose that a photon is a soliton propagating in the dipole
aether with a speed of~$c(U)$, cf. Eq.\,(\ref{eq.local_speed}), controlled by
the dipoles moving in the direction of propagation of the photon.
The dipole distribution thus determines the gravitational index of
refraction, cf. Eq.\,(\ref{eq.refraction}), and consequently the speed of
light~$c(U)$ at the potential~$U$. This solves the problem formulated
in relation to Eq.\,(\ref{eq.local_speed}) and might be relevant for other
phenomena, such as gravitational lensing and the cosmological redshift
\citep[cf., e.g.][]{Ell10}.
Should the speculation in Sect.\,\ref{sss.D_interact} be taken seriously that
the dipole distribution corresponds to
DM, it has to be much more evenly distributed than previously
thought \citep{Hiletal}.
The light deflection would then be caused by gravitationally induced index of
refracion variations.

%%%%%%%%%%%%%%%%%%%%%%%%%%%%%%%%%%%%%%%%%%%%%%%%%%%%%%%%%%%%%%%%%%%%%%%%%%%%%%
%%%%%%%%%%%%%%%%%%%%%%%%%%%%%%%%%%%%%%%%%%%%%%%%%%%%%%%%%%%%%%%%%%%%%%%%%%%%%%
%%%%%%%%%%%%%%%%%%%%%%%%%%%%%%%%%%%%%%%%%%%%%%%%%%%%%%%%%%%%%%%%%%%%%%%%%%%%%%
%%%%%%%%%%%%%%%%%%%%%%%%%%%%%%%%%%%%%%%%%%%%%%%%%%%%%%%%%%%%%%%%%%%%%%%%%%%%%%

\section{Discussion and conclusions}
\label{s.concl}
%% Sect. 4

With Newton's law of gravitation as starting point, the ideas presented
in Sect.\,\ref{ss.Newton}
allow an understanding of far-reaching gravitational force between massive
particles as local interactions of hypothetical massless gravitons
travelling with the speed of light in vacuum. The gravitational attraction
leads to a general mass accretion of massive particles with time, fuelled by
a decrease of the graviton energy density in space.
The physical processes during the conversion of gravitational potential energy
into kinetic energy have been described for two bodies with masses~$m_{\rm A}$
and $m_{\rm b}$ and the source of the potential energy could be identified
in Sect.\,\ref{sss.Gravity}. In order to avoid conflicts with energy and
momentum conservation, we had to modify a
detail of the interaction process in Eq.\,(\ref{eq.G_imbalance}), i.e.,
assume an anti-parallel of the secondary graviton with respect to the
incoming one.

Multiple interactions of gravitons leading to shifts of the effective
gravitational centre of a massive body from the ``centre of gravity''
are treated in Sects. \ref{ss.perihel} to \ref{ss.galaxies} taking
the modified concept into account. The interaction of gravitons with photons
in Sect.\,\ref{ss.Shapiro} had to be modified as well, but
the modification did not change the results, with the exception that now
both the energy and momentum conservation principles are fulfilled.

Our main aim in Sect.\,\ref{ss.redshift} was to identify a physical process
that leads to a speed~$c(U)$ of
photons controlled by the gravitational potential~$U$. This could be achieved
by postulating an aether model with moving dipoles, in which a
gravitational index of refraction $n_{\rm G}(U) = c_0/c(U)$ regulates the
emission and propagation of photons as required by energy and momentum
conservation principles. The emission process thus follows Steps~(1) to (3)
in Sect.~\ref{ss.redshift}, where the local speed of light is given by the
gravitational index of refraction~$n$. In this sense, the statement that an
atom cannot detect the potential~$U$ by \citet{Mueetal} is correct; the local
gravity field~$\vec{g}$, however, is not controlling the emission process.

A photon will be emitted by an atom with appropriate energy and momentum
values, because the local speed of light requires an adjustment of the
momentum. This occurs in the interaction region between the atom and its
environment as outlined in Step~(3).

In the framework of a recently proposed electrostatic impact model in Paper~2,
the physical processes related to the variation of the
electrostatic potential energy of two charged bodies have been described and
the ``source region'' of the potential energy in such a system
could be identified and is summarized in Sect.\,\ref{sss.elec}.

\citet{Sotetal} made a statement in the context of gravitational theories in
`A no-progress report': `[...] it is not only the mathematical formalism
associated with a theory that is important, but the theory must also include a
set of rules to interpret physically the mathematical laws'. With this goal
in mind we have presented our ideas on the gravitational and electrostatic
interactions.

%%%%%%%%%%%%%%%%%%%%%%%%%%%%%%%%%%%%%%%%%%%%%%%%%%%%%%%%%%%%%%%%%%%%%%%%%%%%%%
%%%%%%%%%%%%%%%%%%%%%%%%%%%%%%%%%%%%%%%%%%%%%%%%%%%%%%%%%%%%%%%%%%%%%%%%%%%%%%
%%%%%%%%%%%%%%%%%%%%%%%%%%%%%%%%%%%%%%%%%%%%%%%%%%%%%%%%%%%%%%%%%%%%%%%%%%%%%%
%%%%%%%%%%%%%%%%%%%%%%%%%%%%%%%%%%%%%%%%%%%%%%%%%%%%%%%%%%%%%%%%%%%%%%%%%%%%%%

\acknowledgments
This research has made extensive use of the Smithsonian Astrophysical
Observatory (SAO)/National Aeronautics and Space Administration (NASA)
Astrophysics Data System (ADS).
Administrative support has been provided by the Max-Planck-Institute for
Solar System Research in G\"ottingen, Germany, and the Indian Institute of
Technology (Banaras Hindu University) in Varanasi, India.

%%%%%%%%%%%%%%%%%%%%%%%%%%%%%%%%%%%%%%%%%%%%%%%%%%%%%%%%%%%%%%%%%%%%%%%%%%%%%%
%%%%%%%%%%%%%%%%%%%%%%%%%%%%%%%%%%%%%%%%%%%%%%%%%%%%%%%%%%%%%%%%%%%%%%%%%%%%%%
%%%%%%%%%%%%%%%%%%%%%%%%%%%%%%%%%%%%%%%%%%%%%%%%%%%%%%%%%%%%%%%%%%%%%%%%%%%%%%
%%%%%%%%%%%%%%%%%%%%%%%%%%%%%%%%%%%%%%%%%%%%%%%%%%%%%%%%%%%%%%%%%%%%%%%%%%%%%%

\end{document}